\documentclass[11pt ]{article}
\usepackage{amsmath}
\usepackage{graphicx}
\usepackage{hyperref}
\usepackage{lmodern}
\usepackage{amssymb}
\usepackage{booktabs}
\usepackage{tikz}
\usetikzlibrary{patterns}
\renewcommand{\theequation}
{\arabic{section}.\arabic{equation}}

\makeatletter
\def\eqnarray{ \stepcounter{equation} \let\@currentlabel=\theequation
 \global\@eqnswtrue
 \global\@eqcnt\z@
 \tabskip\@centering
 \let\\=\@eqncr
 $$\halign to \displaywidth\bgroup\@eqnsel\hskip\@centering
 $\displaystyle\tabskip\z@{##}$&\global\@eqcnt\@ne
 \hfil$\displaystyle{{}##{}}$\hfil
 &\global\@eqcnt\tw@$\displaystyle\tabskip\z@{##}$\hfil
 \tabskip\@centering&\llap{##}\tabskip\z@\cr}
\makeatother

\makeatletter
\def\@arrayacol{\edef\@preamble{\@preamble \hskip .5\arraycolsep}}
\def\array{\let\@acol\@arrayacol \let\@classz\@arrayclassz
\let\@classiv\@arrayclassiv \let\\\@arraycr\def\@halignto{}\@tabarray}
\makeatother

\makeatletter
\newcounter{subeqncnt}
\def\thesubeqncnt{\alph{subeqncnt}}
\def\subequations{\begingroup%
   \stepcounter{equation}\edef\@tempa{\theequation}%
   \let\c@equation\c@subeqncnt\c@subeqncnt\z@
   \edef\theequation{\@tempa\noexpand\thesubeqncnt}}

\makeatother

\newcommand{\be}{\begin{equation}}
\newcommand{\ee}{\end{equation}}

\newcommand{\beqa}{\begin{eqnarray}}
\newcommand{\eeqa}{\end{eqnarray}}
\newcommand{\nn}{\nonumber}

\def\CA {{\cal A}}

\def\CF {{\cal F}}
\def\CG {{\cal G}}

\def\CL {{\cal L}}

\def\CN {{\cal N}}
\def\CO {{\cal O}}

\begin{document}

\setlength{\baselineskip}{7mm}
\begin{titlepage}
\begin{flushright}
{\tt NRCPS-HE-52-2021} \\
September, 2021
\end{flushright}

\vspace{1cm}

\begin{center}
{\it \Large     Gauge Field Theory Vacuum \\
and \\ 
Cosmological Inflation without Scalar Field  
 }

\vspace{1cm}

{ {George  Savvidy  }}

\vspace{1cm}

 {\it Institute of Nuclear and Particle Physics, NCSR Demokritos, GR-15310 Athens, Greece}\\
{\it  A.I. Alikhanyan National Science Laboratory, Yerevan, 0036, Armenia}

\end{center}

\vspace{1cm}

\begin{abstract}

 We derive the quantum energy-momentum tensor and the corresponding quantum equation of state for gauge field theory using the effective Lagrangian approach. The  energy-momentum tensor has a term proportional to the space-time metric and provides a finite non-diverging  contribution to the effective cosmological term. This allows to investigate the influence of the gauge field theory vacuum polarisation on the evolution of Friedmann cosmology, inflation  and primordial gravitational waves. The Type I-IV solutions of the Friedmann equations induced by the gauge field theory vacuum polarisation provide an alternative inflationary mechanism and a possibility for late-time  acceleration. The Type II solution of the Friedmann equations generates the initial exponential expansion of the universe of finite duration and the Type IV solution demonstrates late-time acceleration. The solutions fulfil the necessary conditions for the amplification of primordial gravitational waves.

\end{abstract}

\end{titlepage}

\section{\it  Introduction } 
In this article we investigate the influence of the quantum energy-momentum tensor of gauge field theory  resulting from the vacuum polarisation \cite{Savvidy:1977as,Savvidy:2019grj,PhDTheses,Batalin:1976uv,Matinyan:1976mp} and of the corresponding quantum equation of state  on the evolution of Friedmann cosmology \cite{Friedman,Friedman1}, inflation \cite{Starobinsky:1980te,Guth:1980zm,Mukhanov:1981xt,Guth:1982ec,Mukhanov:2005sc,Linde:2007fr,Linde:2015edk,Bezrukov:2007ep,Cervantes-Cota:1995ehs,Wald,Liddle}  and primordial gravitational waves \cite{Lifshitz:1945du,LandauLifshitz,Grishchuk,Grishchuk1,Starobinsky:1979ty,Rubakov:1982df}.

The deep  interrelation  between elementary particle physics and cosmology manifests itself when one considers the contribution of  quantum fluctuations of  vacuum fields to the effective cosmological term  $\Lambda_{cc}$ \cite{Zeldovich:1968ehl,Weinberg,Adler:1982ri,Mukhanov:2005sc,Linde:2007fr,Linde:2015edk,Wald,Liddle,LandauLifshitz,Pasechnik:2016twe}.  In discussing the cosmological constant problem, it is assumed that $\Lambda_{cc}$ corresponds to the vacuum energy density, for which there are many contributions and that anything that contributes to the energy density of the vacuum acts as a cosmological term.  The contribution of zero-point energy exceeds by many orders of magnitude the observational cosmological upper bound on the energy density of the universe.  However the recent covariant calculation of all components of the energy-momentum tensor performed by Donoghue demonstrated that in the case of massless fields the zero-point energy contribution  vanishes   \cite{Donoghue:2020hoh} and  there is no modification of the  cosmological term by the zero-point energy of the massless fields \cite{Donoghue:2020hoh,Akhmedov:2002ts,Ossola:2003ku,Fradkin:1973wke}.

The calculation of the effective Lagrangian in QED by Heisenberg and Euler was the first example of a well-defined physically motivated prescription allowing to obtain a finite, gauge   and renormalisation group-invariant result  when investigating the vacuum fluctuations of quantised fields \cite{Heisenberg:1935qt}.  It appears that only the difference between vacuum energy in the presence and in the absence of external sources has a well-defined physical meaning \cite{Heisenberg:1935qt,Euler:1935zz,Schwinger:1951nm,Coleman:1973jx,Vanyashin:1965ple,Skalozub:1975ab,Brown:1975bc,Duff:1975ue,Savvidy:1977as,Savvidy:2019grj,PhDTheses,Batalin:1976uv,Matinyan:1976mp}. Here we will follow this prescription and will derive the quantum equation of state for non-Abelian gauge fields by using the effective Lagrangian approach \cite{  Nielsen:1978rm,Skalozub:1978fy,Nielsen:1978zg,Ambjorn:1978ff,Nielsen:1978tr,Nielsen:1979xu,Nielsen:1979ta,Nielsen:1979vb,Ambjorn:1979xi,Ambjorn:1980ms,Skalozub:1980nv,Leutwyler:1980ev,Leutwyler:1980ma,Duff:1977ay,parthasarathy,Dittrich:1983ej,Zwanziger:1982na,Flory:1983td,Pagels:1978dd,Mandelstam:1979xd,Taylor:1988vt,Reuter:1994yq} and analyse the properties of Friedmann cosmology driven by the quantum Yang-Mills equation of state.

Let us first review in short  the basic properties of  Friedmann equations  and the standard contributions to the energy density and pressure by dust, radiation and barotropic fluid \cite{LandauLifshitz,Mukhanov:2005sc,Linde:2007fr,Wald,Liddle}.  The equation of state of matter in the universe defines the cosmological evolution and enters on the right-hand side of the first and of the second Friedmann equations  \cite{Friedman, Friedman1}:
\beqa\label{FriedmannEquations}
&&   {k \over a^2}+  {\dot{a}^2 \over a^2} = {8\pi G \over 3 c^4} \epsilon,~~~~~~~   
 {k \over a^2} +  {\dot{a}^2 \over a^2} +2 {\ddot{a} \over a} = - {8\pi G \over c^4} p,  
\eeqa
where $\epsilon$ is the energy density, $p$ is a pressure, and $\dot{a} = {d a / c d t}$. 
The  scale factor $a(t)$ enters into the metric  as \cite{LandauLifshitz,Mukhanov:2005sc,Linde:2007fr,Wald}
\beqa
d s^2 = c^2 d t^2 - a^2(t)  \begin{cases} d\chi^2    +\chi^2 d \Omega^2   \\ d\chi^2 + \sin^2 \chi  d \Omega^2   \\ d\chi^2 + \sinh^2 \chi d \Omega^2   \end{cases}.
\eeqa
These are comoving coordinates; the universe expands or contracts as $a(t)$ increases or decreases, and  the matter coordinates remain fixed. The conformal time $\eta$  is defined as  $c dt = a(\eta) d \eta$. 
It is convenient to transform the Friedmann equations (\ref{FriedmannEquations}) into the following form \cite{LandauLifshitz,Mukhanov:2005sc,Wald}:
\beqa\label{edens}
&& \dot{\epsilon} + 3 {\dot{a} \over a} (\epsilon + p) =0,\\
\label{accel}
&&  {\ddot{a} \over a} = -{4\pi G \over3  c^4} (\epsilon +3 p).  
\eeqa
It follows that the matter equation of state in the universe $p =p(\epsilon)$ defines the behaviour of the solutions of the Friedmann equations. In the case of dust of zero pressure   $p=0, \epsilon =const$ it follows from (\ref{edens})   that
$\epsilon \ a^3 = const$  
and in the case of pure radiation $p= \epsilon/3  $ that
$\epsilon \ a^4 = const.$
For the  general parametrisation of the equation of state $p = w \epsilon$ in terms of the  barotropic   parameter $w$  the solution of (\ref{edens}) has the following form:
\be\label{standradmatter}
 ~~~\epsilon \ a^{3(1+w)} = const,~~~~~
\ee
and when  $w=-1$,  $p = -\epsilon < 0$, it follows from (\ref{accel})   that the acceleration is positive:
$$
{\ddot{a} \over a} = {8\pi G \over 3  c^4} \epsilon > 0.
$$  The equation of state $p = -\epsilon < 0$ is equivalent to having a fluid of positive energy density and negative pressure.  Representation of the dark energy as a barotropic fluid  provides a sufficient condition for the accelerating  expansion of the universe \cite{Guth:1980zm,Mukhanov:1981xt,Guth:1982ec,Peebles:1998qn,Mukhanov:2005sc,Linde:2007fr,Wald,Liddle}.

Most of the studies of inflation are carried out under the general hypothesis that inflation is driven by a scalar field \cite{Mukhanov:2005sc,Linde:2007fr,Bezrukov:2007ep,Kallosh:2013lkr,Ferreira:2021ctx, Antoniadis:2021axu}.  A negative pressure fluid is realised with a scalar field driven inflation   where $\epsilon ={1\over 2}\dot{\phi}^2 +V(\phi)$,~ $p = {1\over 2} \dot{\phi}^2 -V(\phi)$ and   
$\epsilon +p = \dot{\phi}^2 \geq 0,~\epsilon +3p = 2 \dot{\phi}^2 -2 V(\phi).$
The inflationary condition $\epsilon +3p <0$ can be satisfied when the scalar field is in its vacuum state: $V^{'}(\phi_0)=0,~ V(\phi_0) >0, ~\dot{\phi}_0=0$.   It follows that the strong energy dominance condition
$
\epsilon +3p \geq 0
$
 is violated  when~
 $
 p=-\epsilon =-V(\phi_0) < 0
 $ 
and the  energy momentum tensor $T_{\mu\nu}= g_{\mu\nu} V(\phi_0) $ imitates the effective cosmological term in (\ref{accel}):
\be\label{strongenergydomin}
{\ddot{a} \over a} = {8\pi G \over 3  c^4} V(\phi_0) > 0.
\ee
In this paper we will consider the influence of the quantum energy-momentum tensor of gauge field theory \cite{Savvidy:1977as,Savvidy:2019grj,PhDTheses,Batalin:1976uv,Matinyan:1976mp} on the evolution of Friedmann cosmology, inflation and primordial gravitational waves. In Section 2 we will derive the quantum equation of state (\ref{equationofstate1}) for the non-Abelian gauge fields by using the effective Lagrangian approach,  and in Section 3 we will analyse the properties of  Friedmann cosmology driven by the quantum Yang-Mills equation of state (\ref{basiceq}) and (\ref{basiceq1}).  In the subsequent four sections we will demonstrate that the nonsingular Type II solution of the Friedmann equations (\ref{bfactor}),  (\ref{solk-1-II}) provides an alternative mechanism for  a very early stage inflation  of a finite duration (\ref{de-accelerationII}), (\ref{effectivew})  and that there is no initial singularity (\ref{interval-II}). The Type IV solution provides an early-time expansion of the universe that follows a prolongated  phase where the universe remains almost static and subsequently induces a late-time acceleration of a finite duration (\ref{solk-1-IV}). The Type I solution represents the universe that recollapses in a finite time.  The parameters of Type III solution are such that the universe asymptotically approaches a static universe. Infinitesimal deviation of the Type III parameters will place it ether into the Type II or Type IV solutions.   
In Sections 8 and 9 we consider the solutions of the Friedmann  equations in the cases of flat $k=0$  and  positive curvature $k=1$ geometries that appear to represent the evolution of the universe of a finite duration. 

In the last Section 10 we discuss the generation of primordial gravitational waves. The coefficient of amplification $K$ of primordial gravitational waves obtained   in \cite{Grishchuk} 
has the following form:
\be 
K = {1\over 2} \Big(   {\beta \over  n \eta_0}  \Big)^2, ~~~~~\beta = {1-3 w \over 1+3w},\nn
\ee
where $w$ is  the  barotropic  parameter  (\ref{standradmatter}), $n$ is the wave number, the wavelength is $\lambda = 2\pi a/ n$ and the wave had been initiated at conformal time $\eta_0$. A gravitational wave is amplified when equation of state differs from that of radiation ($p = {1\over 3}\epsilon$, $w= 1/3$).  

In the case of quantum gauge field theory the equation of state has the following form (\ref{energypressure3}):
\be  
 \epsilon =      {  \CA \over \tilde{a}^4(t)} \Big(\log{ {1 \over \tilde{a}^4(t)  }} -1\Big) \Lambda^4_{YM} ,~~~~~~~p =     { \CA \over 3 \tilde{a}^4(t)}   \Big(\log{ {1 \over \tilde{a}^4(t)  }} +3\Big) \Lambda^4_{YM} ,
 \ee 
where $a = a_0 \tilde{a}$  (\ref{dimless}) and $a_0$ is the initial data parameter (\ref{fieldstrength1}).   The relations between  energy density,  pressure and the effective  parameter $w$ have the following form (\ref{relation}):
 \be 
 p = {1\over 3}\epsilon + {4\over 3}  {\CA \over \tilde{a}^4(t)} \Lambda^4_{YM}, ~~~~~~~~~ w = {p\over \epsilon} ={    \log{ {1 \over \tilde{a}^4(t)  }} +3    \over  3 \Big(\log{ {1 \over \tilde{a}^4(t)  }} -1\Big)  }.
 \ee
The first relation deviates from  $p = {1\over 3}\epsilon$ by the additional term depending on the  coefficient $\CA$ (\ref{mattergroupparameter}), the scale $\Lambda_{YM}$ and the scale factor $\tilde{a}(t)$.  The deviation is large at the initial stages of  expansion of the universe and tends to zero at a late time when  $\tilde{a}(t) \rightarrow \infty$.  Therefore the tensor perturbation of the Type II and Type IV cosmologies naturally amplify the primordial gravitational waves.

\section{\it  Quantum Yang-Mills Equation of State }

 We will assume here that the universe has in it only fluctuating vacuum  gauge fields and will neglect the contributions to the energy density from radiation,  elementary particles of the Standard Model or of the Grand Unified  Theory (GUT).  The contribution of the radiation and of other matter components can be added afterwards.  We will derive the equation of state by using the explicit expression for the effective Lagrangian in the Yang-Mills gauge field theory \cite{Savvidy:1977as,Savvidy:2019grj,PhDTheses,Batalin:1976uv,Matinyan:1976mp}. The effective Lagrangian is a sum of the Heisenberg-Euler Lagrangian $ \CL_q$ \cite{Heisenberg:1935qt}   taken in the limit of massless chiral fermions \cite{Savvidy:2019grj}:
\beqa\label{chirallimit}
 \CL_q&=& -\CF + {    N_f  \over 48 \pi^2}  g^2 \CF
\Big[  \ln ({2 g^2 \CF \over \mu^4})  - 1  \Big] ~,~
\eeqa
where $N_f$ is the number of fermion flavours and of the Yang-Mills effective Lagrangian $\CL_g$ for SU(N) gauge field theory \cite{Savvidy:1977as,Savvidy:2019grj,PhDTheses}:
\be\label{YMeffective0}
\CL_g  =  
-\CF - {11  N \over 96 \pi^2} g^2 \CF \Big( \ln {2 g^2 \CF \over \mu^4}- 1\Big)~,~~~~~~~~~~~
\CG =  G^*_{\mu\nu}G^{\mu\nu} =0~,
\ee 
where  the vacuum field $\CF =    {1\over 4} g^{\alpha\beta} g^{\gamma\delta}G^a_{\alpha\gamma}G_{\beta\delta}  \geq 0$ is of a chromomagnetic  nature.   The one-loop effective Lagrangian has  exact logarithmic dependence on  the invariant $\CF $.  The effective Lagrangian allows to obtain  the quantum energy momentum tensor $T_{\mu\nu}$  by using the  expressions (\ref{chirallimit})  and (\ref{YMeffective0}) \cite{Savvidy:2019grj}: 
\beqa\label{energymomentumYM01}
T_{\mu\nu} = T^{YM}_{\mu\nu}\Big[1 +{ b \ g^2 \over 96 \pi^2} 
  \ln {2 g^2 \CF \over \mu^4} \Big]
- g_{\mu\nu}    { b \   g^2 \over 96 \pi^2}  \CF  ,~~~~~~~~~\CG=0,
\eeqa
where  $  b =11N - 2N_f$. The vacuum energy density  has therefore the following form:
\be\label{energyexpYM0}
T_{00} \equiv   \epsilon(\CF)=  ~\CF + {b\ g^2\over 96 \pi^2}  \CF \Big( \ln {2 g^2 \CF \over \mu^4}- 1\Big)
\ee
and the spacial components of the stress tensor are:
\beqa\label{stresstensorYM0}
T_{ij}  =  \delta_{ij} ~\Big[ {1\over 3}   \CF + {1\over 3}  {b\ g^2\over 96 \pi^2}  
 \CF \Big( \ln {2 g^2 \CF \over \mu^4}+3 \Big) \Big] = \delta_{ij}  ~p (\CF).
\eeqa
Thus we have the following quantum gauge field theory equation of state:
\beqa\label{equationofstate}
\epsilon(\CF)= ~\CF + {b\ g^2\over 96 \pi^2}  \CF \Big( \ln {2 g^2 \CF \over \mu^4}- 1\Big),~~~~~~~~~
p (\CF)=  {1\over 3}   \CF + {1\over 3}  {b\ g^2\over 96 \pi^2}  
 \CF \Big( \ln {2 g^2 \CF \over \mu^4}+3 \Big). 
\eeqa 
 \begin{figure}
 \centering
\includegraphics[angle=0,width=7cm]{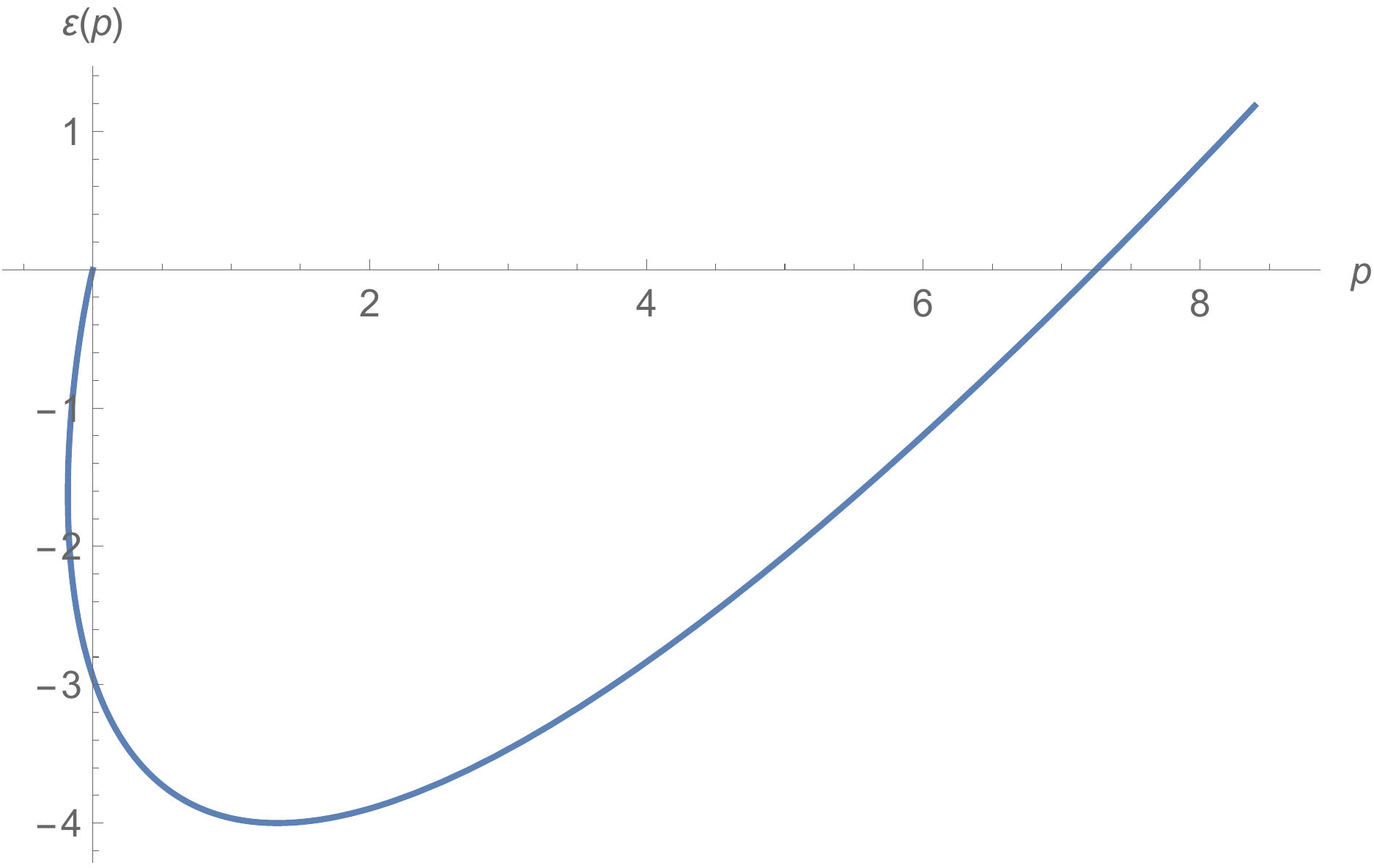}
\centering
\caption{There are regions in the phase space $(\epsilon,p)$ of  the quantum Yang-Mills states (\ref{equationofstate1})   where $\epsilon$  and $p$ are positive, where $p$ is positive and $\epsilon$ is negative and where they are both negative. }
\label{fig3} 
\end{figure}
The energy density $\epsilon(\CF)$ has its minimum outside of the perturbative vacuum  state $  \CF  =0$ at the Lorentz and renormalisation group invariant field strength \cite{Savvidy:1977as}
\be\label{chomomagneticcondensate0}
 2 g^2 \CF_{vac}=    \mu^4  \exp{(-{96 \pi^2 \over b\ g^2(\mu) })}= \Lambda^4_{YM}, 
\ee
which characterises the dynamical breaking of  scaling invariance  of  YM theory (\ref{energymomentumYM01}):
\be
T^{\mu}_{\mu}=  - { b    \over 48 \pi^2}   2 g^2 \CF_{vac} .\nn
\ee 
Thus the equation of state (\ref{equationofstate}) will take the following form:
\beqa\label{equationofstate1}
\epsilon(\CF)  =~  {b\ g^2\over 96 \pi^2}  \CF \Big( \ln {2 g^2 \CF \over \Lambda^4_{YM}}- 1\Big),~~~~~~~~~~
p (\CF)=  ~{1\over 3}  {b\ g^2\over 96 \pi^2}  
 \CF \Big( \ln {2 g^2 \CF \over \Lambda^4_{YM}}+3 \Big) .
\eeqa 
By expressing the vacuum field strength tensor $\CF$ in terms of vacuum pressure  $\CF = \CF(p)$ and substituting it into the vacuum energy density we will get the equation of state in the form $\epsilon = \epsilon(p)$  shown in Fig.\ref{fig3}. 
In the limit  $2 g^2 \CF \gg \Lambda^4_{YM}$  (\ref{equationofstate1})   reduces to a radiation equation of state:   
$
p= \epsilon/3. 
$
There are regions in the phase space of states $(\epsilon,p)$ where $\epsilon$  and $p$ are positive, where $p$ is positive and $\epsilon$ is negative and where they are both negative, as it is shown in  Fig. \ref{fig3}. The pressure is always higher than in the case of radiation equation of state: 
\be
p= {1 \over 3} \epsilon + {4 \over 3}{b\ g^2 \CF \over 96 \pi^2} \Lambda^4_{YM}~~~~~ \text{and}~~~~~w= {p\over \epsilon} = {\ln {2 g^2 \CF \over \Lambda^4_{YM}}+3 \over 3 \Big( \ln {2 g^2 \CF \over \Lambda^4_{YM}}-1 \Big)}
\ee 
It also follows from the  energy momentum-tensor expression (\ref{energymomentumYM01}) that when the gauge field is in its ground state (\ref{chomomagneticcondensate0}),  $T^{\mu\nu}$ is proportional to the space-time metric  $g^{\mu\nu} $:
\be\label{vacuumenergyden0}
 T^{\mu\nu}_{vac} = - g^{\mu\nu}   ~ { b    \over 192 \pi^2}  2  g^2 \CF_{vac} ,
\ee
and  equation of state reduces to the equation $ p=- \epsilon  > 0 $.  The equation of state $p = -\epsilon > 0$ is equivalent to having a fluid of positive pressure  and negative energy density alternative to the  inflation that is driven by a scalar field (\ref{strongenergydomin}).

In the next sections we will analyse the Friedmann cosmology that is  driven by the vacuum gauge field theory  equation of state (\ref{equationofstate1}). The Einstein equation  in the presence of the  vacuum energy momentum tensor  (\ref{energymomentumYM01}) has the following form\footnote{The r.h.s in (\ref{HAequation}) is the contribution of the  gauge field theory vacuum polarisation  $<T^{YM}_{\mu\nu}>$ (\ref{energymomentumYM01}). There are no particles in the initial state of the universe.   
The effect is similar to the Starobinsky   vacuum polarisation by the gravitational field leading to the contribution $<T^{GR}_{\mu\nu}>  \sim f(R^2)$  \cite{Starobinsky:1980te}.}: 
\be\label{HAequation}
R_{\mu\nu} - {1\over 2} g_{\mu\nu} R =    {8\pi G \over c^4}  \Big[ T^{YM}_{\mu\nu}\Big(1 +{ b \ g^2 \over 96 \pi^2} 
  \ln {2 g^2 \CF \over \mu^4} \Big)
- g_{\mu\nu}    { b \   g^2 \over 96 \pi^2}  \CF \Big].
\ee
It follows that the induced effective cosmological term can be  expressed in terms of vacuum energy density (\ref{equationofstate1}) and vacuum field (\ref{chomomagneticcondensate0}) as
\be\label{gap}
 \Lambda_{eff}    = {8 \pi G  \over 3 c^4} ~\epsilon_{vac}  =-{8 \pi G  \over 3 c^4} { b  \over 192 \pi^2}   2 g^2 \CF_{vac}= -{8 \pi G  \over 3 c^4}  { b   \over 192 \pi^2} \Lambda^4_{YM}  ~.
\ee
During the cosmological evolution the field strength tensor $\CF$ will not stay constantly in its ground state  (\ref{chomomagneticcondensate0}) but will roll through the well-defined trajectory in the phase space  of states $(\epsilon,p)$, which is  defined by the Friedmann equations (\ref{FriedmannEquations}) and (\ref{edens}), (\ref{accel}).   

In general relativity there is no covariantly constant gauge fields and the time evolution of the gauge field is  described by the Yang-Mills equation in the background gravitational field or equivalently can be defined through the covariant conservation of the energy-momentum tensor: 
$
T_{\mu\nu}^{~~;\nu}=0.
$ 
It is the last option we will use in the next section in solving the Friedmann equations. The vacuum space homogeneous  time-dependent  solutions of the Yang-Mills equations were first considered in \cite{Baseyan, Natalia,Asatrian:1982ga,SavvidyPlanes} and in the context of the cosmological models in   \cite{Cervero:1978db,Ford:1989me,Henneaux:1982vs,Hosotani:1984wj,Galtsov:1991un,Gibbons:1993pq,Golovnev:2008cf,Bamba:2008xa,Maleknejad:2011sq,Elizalde:2012yk,Elizalde:2012yk, Adshead:2012qe}.

\section{\it   Quantum Yang-Mills  Equation of State and  Friedmann Cosmology} 

The time derivative of the energy density given in  (\ref{equationofstate1})  is
\be
\dot{\epsilon} =   \CA ~(2 g^2 \dot{\CF})  ~\log{2 g^2 \CF \over \Lambda^4_{YM}},
\ee
where $\dot{\CF}=  {d \CF / c d t}$. The time evolution of the energy density $\epsilon $ in (\ref{edens}) depends on the sign of the sum $\epsilon +p$. By using the expressions for $\epsilon $ and $p$ in (\ref{equationofstate1})  for the sum $\epsilon +p$ we will obtain:
\beqa
\epsilon +p = {4 \CA \over 3} ~  (2 g^2 \CF)~  \log{2 g^2 \CF \over \Lambda^4_{YM}}, 
\eeqa
where $\CA$ is the coefficient of the one-loop $\beta(g)$ function: 
\be\label{mattergroupparameter}
\CA = {b \over 192 \pi^2}={11 N-2 N_f \over 192 \pi^2} .
\ee
It follows that for $2 g^2 \CF < \Lambda^4_{YM} $ the weak energy dominance condition $\epsilon +p \geq 0$ is violated. The equation (\ref{edens}) now takes the form 
\be
2 g^2 \dot{\CF} +4 (2 g^2 \CF)  {\dot{a} \over a}=0
\ee
and can be integrated yielding 
\be\label{fieldstrength1}
2 g^2 \CF~ a^4  =const \equiv    \Lambda^4_{YM}\ a^4_{0}, 
\ee
where the integration constant is parametrised in terms of the {\it initial data parameter} $a_0$. The energy density and pressure (\ref{equationofstate1})  can now be expressed in terms of the scale factor $a(t)$:
\be\label{energymomentum2} 
 \epsilon =      \CA  {a^{4}_{0} \over a^4} \Big(\log{ {a^{4}_{0} \over a^4  }} -1\Big) \Lambda^4_{YM},~~~~~
 p= \CA  {a^{4}_{0} \over 3 a^4} \Big(\log{ {a^{4}_{0} \over a^4  }} +3\Big) \Lambda^4_{YM}.
\ee
With the help of the last expression for the $ \epsilon$ the  first Friedmann  equation (\ref{FriedmannEquations})  will take the following form:
\beqa\label{FriedmannEquations2}
{d a \over c dt } = \pm \sqrt{   {8\pi G \over 3 c^4}~ \CA ~ \Lambda^4_{YM} ~{a^{4}_{0} \over a^2} \Big(\log{ {a^{4}_{0} \over a^4  }} -1\Big)  -k},~~~~~~~~k=0,\pm 1.
\eeqa
It is convenient to define the length scale $L$ as it appears naturally in (\ref{gap}) and (\ref{FriedmannEquations2}):
\beqa\label{basicpar}
&&   {1\over L^2} =  { 8 \pi G  \over 3 c^4}\ \CA\  \Lambda^4_{YM}  \equiv    \Lambda_{eff}  ~ ,
\eeqa
so the equation (\ref{FriedmannEquations2}) will take the following form:
\beqa\label{FriedmannEquations3}
{d a \over c d t} = \pm \sqrt{   {a^{2}_{0} \over L^{2}} ~{a^{2}_{0} \over a^2} \Big(\log{ {a^{4}_{0} \over a^4 }} -1\Big) -k}.
\eeqa
In order to  simplify the evolution equations further  it is convenient  to introduce the dimensionless scale factor  $\tilde{a}$ and the dimensionless time variable $\tau$:
\be\label{dimless}
a(\tau) = a_0 ~\tilde{a}(\tau),~~~~c t = L ~ \tau,~ 
\ee
where we normalise the scale factor $a(\tau) $ to the constant parameter $a_0$ in (\ref{fieldstrength1}).  
In these variables the evolution equation (\ref{FriedmannEquations3}) is in its final form:   
\beqa\label{basiceq}
{d\tilde{a} \over d \tau }= \pm \sqrt{   {1 \over \tilde{a}^2} \Big(\log{ {1 \over \tilde{a}^4 }} -1\Big) -k \gamma^2 },~~~~~~k=0,\pm 1,~~~~~~~~  \gamma^2 =\Big({L \over a_0}\Big)^2 .
\eeqa
The  evolution equation (\ref{basiceq}) can be represented  in terms of the dimensionless conformal time $\eta$:
\be\label{conformaltime}
c dt =  L ~ d \tau = a(\eta) d \eta= a_0 \tilde{a} d \eta,
\ee
 as well as (the prime denotes the differentiation with respect to $\eta$):
\beqa\label{basiceq1}
\tilde{a}^{'} \equiv {d \tilde{a} \over d \eta}= \pm \sqrt{  {1\over \gamma^2} \Big( \log{ {1 \over \tilde{a}^4 } } -1\Big) - k ~ \tilde{a}^2}.
\eeqa
The evolution equations (\ref{basiceq}) and (\ref{basiceq1}) should be investigated in six  regions of the two-dimensional  parameter space   $(a_0,\Lambda_{YM})$. The numerical value of  $\gamma^2$  defines the relation $a^2_0={1\over \gamma^2} L^2(\Lambda_{YM}) $ between basic independent  parameters $a_0$ and $\Lambda_{YM}$ through the equations (\ref{basiceq}) and (\ref{basicpar}).  Thus the corresponding six regions in the parameter space are defined in terms of $\gamma^2$:
\beqa\label{sixregions}
&&k=-1, ~~~~ 0 \leq  \gamma^2 < \gamma^2_c~~~~~~~~~~~~~~ \text{Regions I ($\tilde{a} \leq \mu_1$) and II ($ \mu_2   \leq  \tilde{a}  $)}  \nn\\
&&k=-1, ~~~~ \gamma^2= \gamma^2_c ={2\over \sqrt{e}}~~~~~~~~~~~\text{Region  III (separatrix, $\tilde{a} \leq \mu_c$)} \nn\\
&&k=-1, ~~~~ \gamma^2_c <  \gamma^2 ~~~~~~~~~~~~~~~~~~~~ \text{Regions IV ($0 \leq \tilde{a}   $)}  \\
&&k=0,~~~~~~~ \nn\\
&&k=1,~~~~~~~0 \leq  \gamma^2. \nn
\eeqa
In terms of scale factor  $\tilde{a}$ and time variable  $\tau$ (\ref{dimless}) the field strength  tensor (\ref{fieldstrength1})  has the following form: 
\be\label{fieldstrength}
2 g^2 \CF~ =   {\Lambda^4_{YM}\over \tilde{a}^4(\tau) }
\ee
and the energy density and the pressure (\ref {energymomentum2}) will take the form 
\be\label{energypressure3} 
 \epsilon =      {  \CA \over \tilde{a}^4(\tau)} \Big(\log{ {1 \over \tilde{a}^4(\tau)  }} -1\Big) \Lambda^4_{YM} ,~~~~~~~p =     { \CA \over 3 \tilde{a}^4(\tau)}   \Big(\log{ {1 \over \tilde{a}^4(\tau)  }} +3\Big) \Lambda^4_{YM}. 
 \ee
There is a straightforward relation between  energy density,  pressure and the  barotropic   parameter $w$:
 \be\label{relation}
 p = {1\over 3}\epsilon + {4\over 3}  {\CA \over \tilde{a}^4(\tau)} \Lambda^4_{YM}, ~~~~~~~~~ w = {p\over \epsilon} ={    \log{ {1 \over \tilde{a}^4(\tau)  }} +3    \over  3 \Big(\log{ {1 \over \tilde{a}^4(\tau)  }} -1\Big)  }.
 \ee
 In the next sections we will investigate the solutions of the equation (\ref{basiceq}) and the time evolution of the field strength tensor (\ref{fieldstrength}), of the energy density and the pressure (\ref{energypressure3} ).   We can also extract the Hubble parameter from (\ref{FriedmannEquations}) by using (\ref{basiceq})
\be\label{habble}
L^2 H^2 = L^2 \Big({\dot{a} \over a  }\Big)^2 ={1 \over   \tilde{a}^2} \Big({d\tilde{a} \over d \tau }\Big)^2
 =  {1 \over \tilde{a}^4(\tau)} \Big(\log{1 \over \tilde{a}^4(\tau)} -1\Big) - {k \gamma^2 \over \tilde{a}^2(\tau)}  
\ee
and the corresponding deceleration parameter 
\be
q= -{\ddot{a} \over a} {1\over  H^2}. 
\ee
The acceleration is determined by the right-hand side of the equation (\ref{accel}) and is proportional to $\epsilon +3p$, which is:
\beqa
\epsilon +3 p = 2 \CA  ~  (2 g^2 \CF)~ \Big( \log{2 g^2 \CF \over \Lambda^4_{YM}} +1\Big). 
\eeqa
Similar to the case of the scalar field driven evolution (\ref{strongenergydomin})  here as well for the fields $2 g^2 \CF <  {1\over e} \Lambda^4_{YM} $ the strong energy dominance condition $\epsilon +3p \geq 0$ is violated.  
From acceleration Friedmann equation (\ref{accel}) and (\ref{energypressure3} ) we have 
\be
L^2 {\ddot{a} \over a}  
=-   {1 \over \tilde{a}^4} \Big(\log{1 \over \tilde{a}^4} +1\Big).
\ee
Thus  for $q$ with the help of (\ref{habble}) we will get 
\be\label{deceleration}
q= {{1 \over \tilde{a}^4} \Big(\log{1 \over \tilde{a}^4} +1\Big)   \over {1 \over \tilde{a}^4} \Big(\log{1 \over \tilde{a}^4} -1\Big) - {k \gamma^2 \over \tilde{a}^2}   } 
\ee
and for the density parameter  $\Omega_{vac}$ the following expression:
\be\label{omega}
\Omega_{vac}  \equiv {8 \pi G \over 3 c^4} {\epsilon \over H^2} ={1 \over L^2 H^2} ~  {1 \over \tilde{a}^4} \Big(\log{1 \over \tilde{a}^4} -1\Big),
\ee
where we used  (\ref{energypressure3} ), (\ref{basicpar}). By using the equation  (\ref{habble}) $\Omega_{vac}$ can be expressed also in the following form:
\be\label{omegavac}
\Omega_{vac} -1 =  k { \gamma^2 \over  L^2 H^2 \tilde{a}^2} =  k { \gamma^2 \over  ({d \tilde{a} \over d \tau})^2}. 
\ee
We will  investigate these observables  in the two-dimensional parameter space $(a_0,\Lambda_{YM})$ in each of the six regions (\ref{sixregions}). As we mentioned  above, the parameter $\gamma^2 = {L^2 \over a^2_0}$ is a function of $a_0$ and $\Lambda_{YM}$,  the basic parameters defining the evolution of the Friedmann equations in the case of gauge field theory vacuum polarisation.  We will start our analysis by considering  the $k=-1$ geometry.

\section{\it  The Parameter Space of the Type I-IV Solutions  }

\begin{figure}
 \centering
\includegraphics[angle=0,width=5cm]{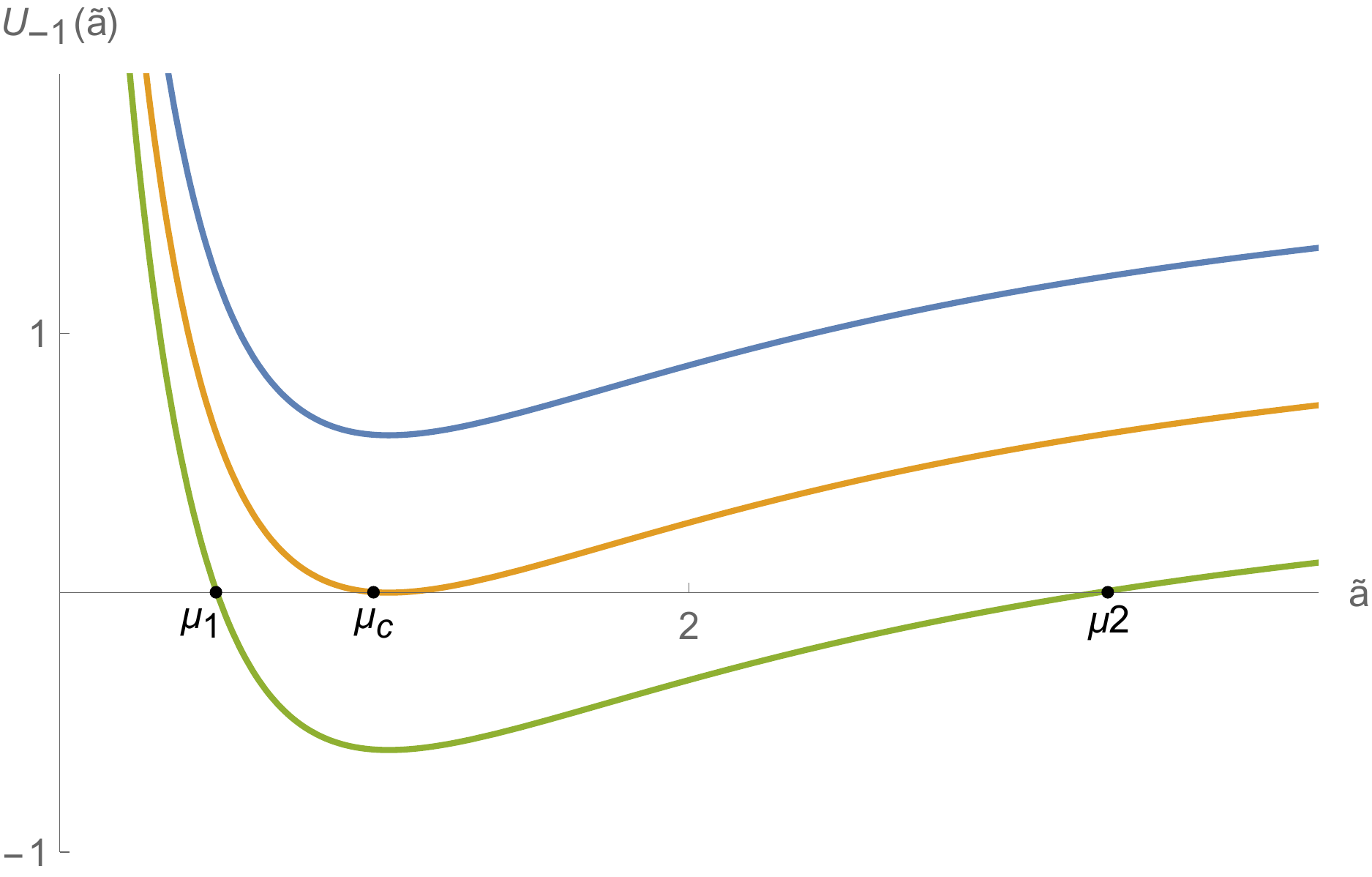}~~~~~
\includegraphics[angle=0,width=5cm]{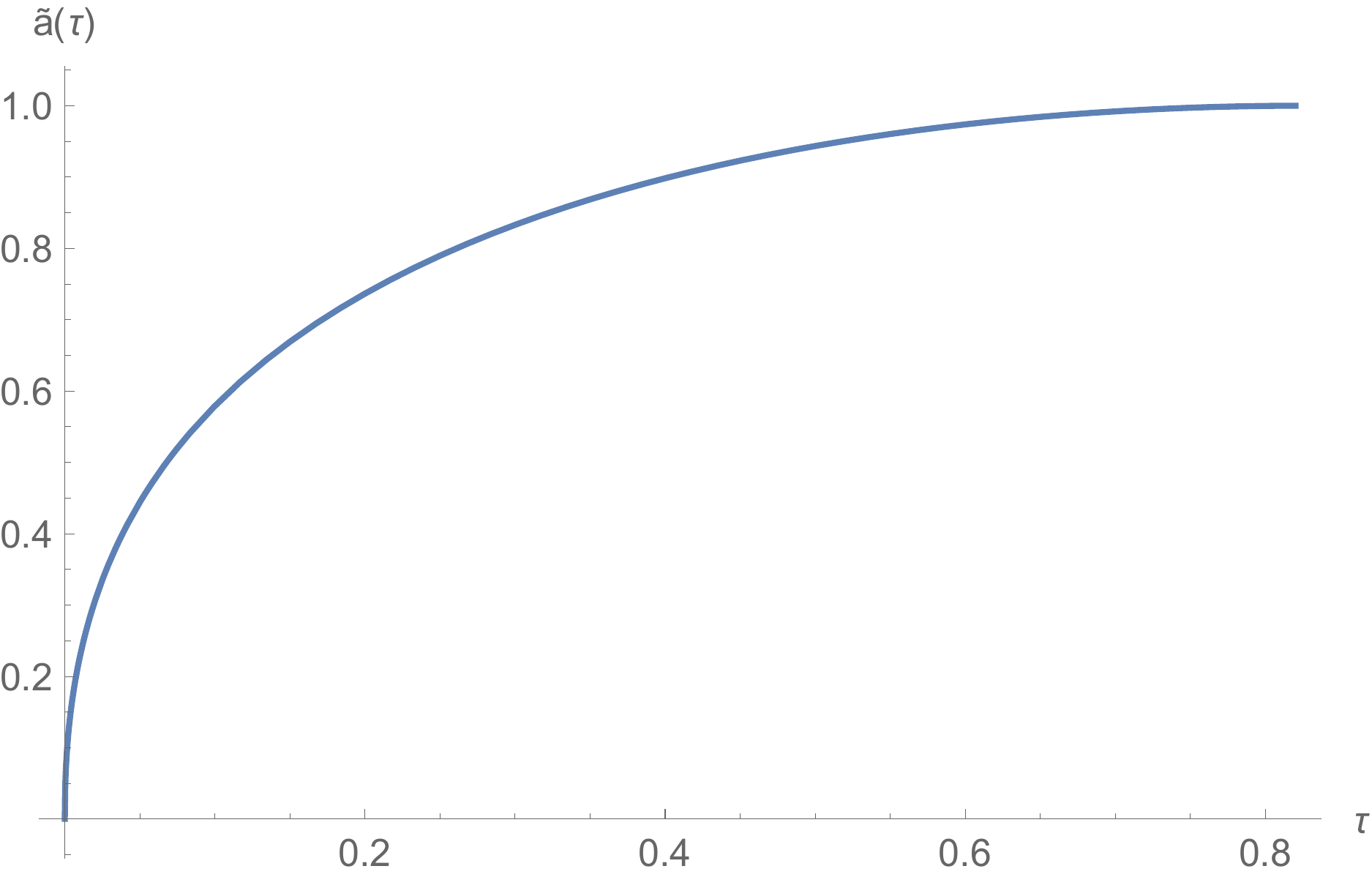}
\includegraphics[angle=0,width=5cm]{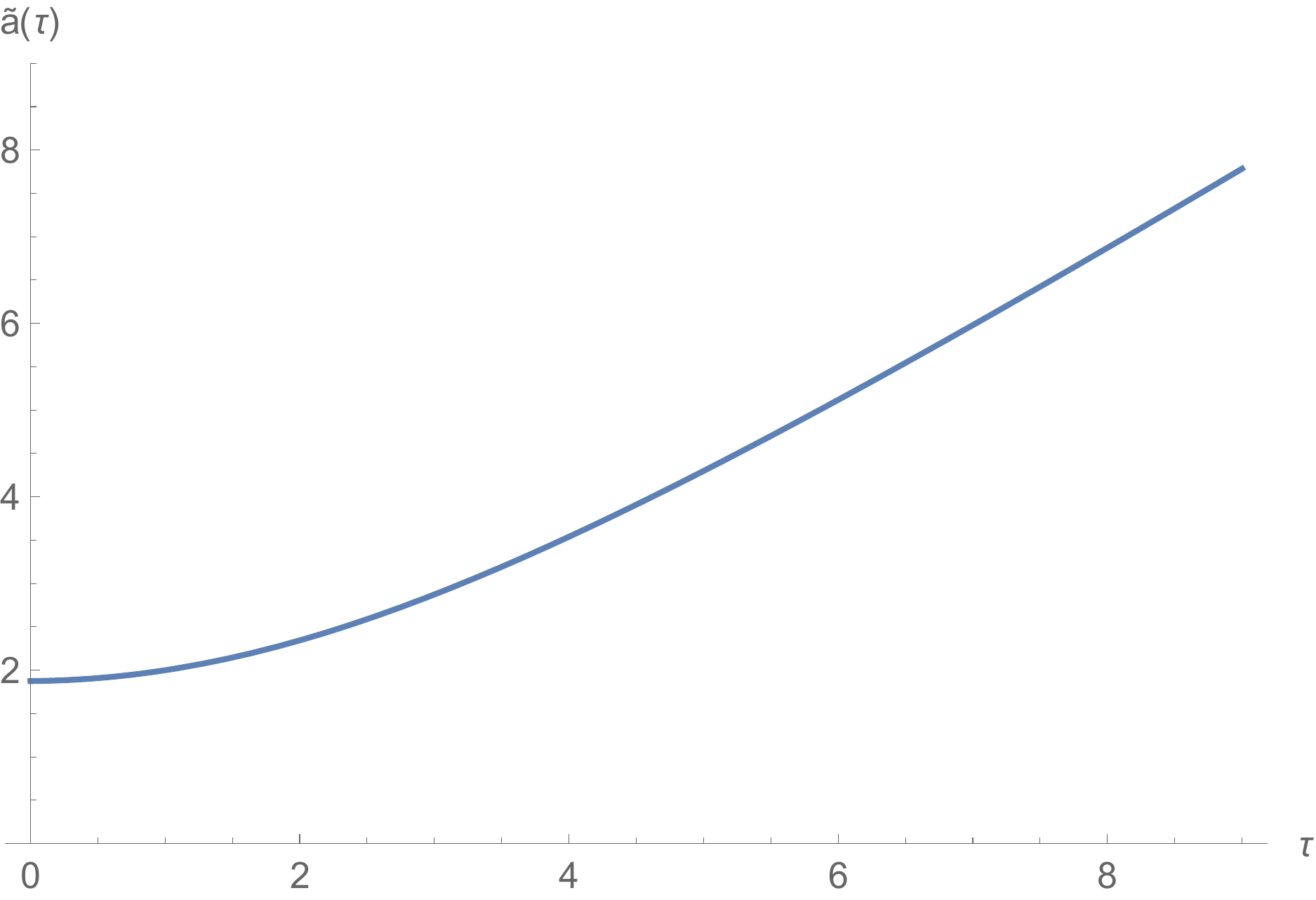}
\centering
\caption{The behaviour of the potential $U_{-1}(\tilde{a})$ (\ref{potenk-1}) is shown in the left figure.  When the parameter $\gamma^2$ is in the interval $0 \leq \gamma^2  < {2\over \sqrt{e}}$, there are two solutions of the equation $U_{-1}(\mu_i)=0,~ i=1,2$, that define the   Region I,  where $\tilde{a} \in [0, \mu_1]$ and the Region II,  where $\tilde{a} \in [\mu_2,\infty]$.   ~When $\gamma^2  =\gamma^2_c = {2\over \sqrt{e}}$, there is only one solution of the equation $U_{-1}(\mu_c)=0$ that defines  the Region III, where $\tilde{a} \in [0,\mu_c]$.  When $ {2\over \sqrt{e}} < \gamma^2  $, the potential is always positive $U_{-1} (\tilde{a}) > 0$  and the Region IV  is where  $\tilde{a} \in [0,\infty]$.   
In particular, when  $\gamma^2 =1$,  $\mu_1 \simeq 1$  and the scale factor $\tilde{a}(\tau)$ of the Type I solution is  bounded $\tilde{a} \in [0, \mu_1]$.  Its  evolution time is  finite $\tau \in [0,2 \tau_m ] $, where $ \tau_m \simeq   0.83$ is a half period.  The figure in the middle  shows the behaviour of the Type I solution.   The Type II solution shown in the right figure is  unbounded  $\tilde{a} \in [\mu_2,\infty]$, where $\mu_2 \simeq 1.87$ and $\tau \in [0,\infty]$. The  Type II solution initially grows exponentially because the deceleration parameter is negative, $q < 0$.  At late time the regime of exponential expansion continuously transforms into a linear in time growth of the scale factor. }
\label{figU-1} 
\end{figure}
In case of $k=-1$ geometry the  equation (\ref{basiceq})  takes the following form:
\beqa\label{k2case}
&&{d \tilde{a} \over d \tau}= \pm \sqrt{ {1 \over \tilde{a}^2} \Big( \log{ {1 \over \tilde{a}^4 } } -1\Big) +  \gamma^2  },~~~~\text{where} ~~~~~0  \leq \gamma^2,
\eeqa
and   the corresponding "potential"  function $U_{-1}(\tilde{a})$ shown in Fig.\ref{figU-1} is:
\be\label{potenk-1}
U_{-1}(\tilde{a}) \equiv   {1 \over \tilde{a}^2}  
\Big( \log{ {1 \over \tilde{a}^4 } } -1\Big) + \gamma^2 .
\ee
The solution of the equation $U_{-1}(\mu) =0$ determines the values of the scale factor $\tilde{a}=\mu$ at which the square root  changes its sign.  The evolution equation (\ref{k2case})  should be restricted to those  real values of $\tilde{a}$ at which the potential $U_{0}(\tilde{a}) $ is nonnegative. Thus the equation $U_{-1}(\mu)=0$  defines  the boundary values of the scale factor $\tilde{a}= \mu $: 
\be\label{k-1Isol}
 {1 \over \mu^2}  
 \Big( \log{ {1 \over \mu^4 } } -1\Big) + \gamma^2=0.~~~~
  \ee
The behaviours of the solutions depending  on the value of the parameter $\gamma^2$. When 
\be\label{TypeIcond} 
0 \leq \gamma^2  <  \gamma^2_c  \equiv {2\over \sqrt{e}} ,
\ee
 there are two solutions $\tilde{a}_1 = \mu_1$ and $\tilde{a}_2 = \mu_2$ of the above equation that are defining the regions where the potential $U_{-1}(\tilde{a})$ is positive. In the first region $I$   we have $\tilde{a} \in [0, \mu_1]$, and   in the second region $II$   $\tilde{a} \in [\mu_2, \infty]$. These two regions are shown in Fig.\ref{figU-1}. The region $III$ appears when $ \gamma^2  = \gamma^2_c   $ and  it is the separatrix  between regions $II$ and $IV$.  At the separatrix  point $ \gamma^2  = \gamma^2_c   $ the equation $U_{-1}(\mu)=0$ has only one solution  $\tilde{a} = \mu_c$ and the scale factor $\tilde{a}$ takes its values in the maximally available  interval $\tilde{a} \in [0, \mu_c]$.   Finally, in the region $IV$, where  $\gamma^2_c < \gamma^2   $, the potential function $U_{-1}(\tilde{a})$ is always positive for all values of $\tilde{a}$ and the scale factor takes its values in the whole interval  $\tilde{a} \in [0, \infty]$. We will consider  these four regions separately. 

\section{\it Type I Solution  }
 
Let's consider first the Type $I$ solution when  $0 \leq \gamma^2  < \gamma^2_c $ and $\tilde{a} \leq \mu_1$. The equation (\ref{k-1Isol}) can be solved by the substitution 
\be\label{mu-u-eq}
 \gamma^2 \mu^2 = 2 u
\ee
that reduces the equation (\ref{k-1Isol}) to the Lamber-Euler type \cite{Lambert,Euler,Corless}: 
\be\label{Lamber-Euler}
~~~~u e^{-u} = {\gamma^2 \over 2 \sqrt{e}}.
\ee
The solution is expressible in terms of $W_0(x)$ function which is defined  for negative values of its argument in the interval $ -1/e < x \leq 0$  and is acquiring negative values   $ -1 < W_0(x)  \leq 0$.  The solution takes the following form (see Appendix A):
\be
u = -W_{0}\Big(-{\gamma^2 \over 2 \sqrt{e}} \Big).
\ee
The maximal value of the scale factor $\tilde{a}$ therefore is
\be\label{solk-1-I}
\mu^2_1 = -{2\over \gamma^2} ~W_{0}\Big(-{\gamma^2 \over 2 \sqrt{e}}\Big),~~~~
\ee
and it follows that (see Appendix)
\be\label{limsI}  
{1\over \sqrt{e} } \leq \mu^{2}_{1} < \sqrt{e},~~~~~~   \gamma^2 \mu^2_1 < 2 .
\ee
The interval in which $ \tilde{a}$ takes its values is:   
\be
~~~\tilde{a} \in [0, \mu_1].
\ee
With the next substitution  
\be
   \tilde{a}^4  =\mu^4_1 e^{-b^2},~~~~  ~ b \in  [-\infty,\infty],
\ee
the equation (\ref{k2case})  will reduce to the following form:  
\be
{d b \over d \tau}  =   {2 \over \mu^2_1}  ~   e^{{b^2 \over 2}}   \Big(1    -     {\gamma^2  \mu^2_1 \over b^2}  (1-  e^{-{b^2 \over 2}}   \Big)^{1/2}.
\ee
With the boundary condition $b(0)=-\infty$  at $\tau =0$ we will get the integral representation of the function $b(\tau)$:
\beqa\label{solk-1}
&&\int^{b(\tau)}_{-\infty}   { d b   e^{-{b^2 \over 2}}   \over 
  \Big(1    -      { \gamma^2  \mu^2_1 \over b^2}  (1-  e^{-{b^2 \over 2}}   \Big)^{1/2}        } ~ = {2 \over \mu^2_1}   ~ \tau.
\eeqa
Within a finite-time interval after the initial expansion the universe will approach its maximal size $\mu_1$ and then begin to recontract. This time interval $\tau \in [0,\tau_m]$ is defined by the integral 
\beqa\label{bondk-1I}
&&\int^{0}_{-\infty}    { d b   e^{-{b^2 \over 2}}   \over 
  \Big(1    -       { \gamma^2  \mu^2_1  \over b^2}  (1-  e^{-{b^2 \over 2}}   \Big)^{1/2}        } ~ = {2 \over \mu^2_1}   ~ \tau_m 
\eeqa
and is equal to the half of the total period, which is equal to $2\tau_m$.  Thus during the time evolution $\tilde{a}(\tau)$ is reaching its maximal value $\tilde{a}=\mu_1$ at $\tau_m$ and then contracting  to zero value at $\tau=2 \tau_m$. The asymptotic behaviour of  the $\tilde{a}(\tau)$ near $\tau=\tau_m$ and $\tau =0$  has the form 
\beqa\label{singu}
\tilde{a}(\tau) \simeq   \begin{cases}  \mu_1 -  {2-\gamma^2  \mu^2_1 \over 2 \mu^3_1}  (\tau-\tau_m)^2,~~~~~ \tau \rightarrow \tau_m \\   \tau^{1/2} \ln^{1/4}{1\over \tau^{1/2}},~~~~~~~~~~~~~\tau \rightarrow 0   \end{cases}. 
\eeqa
At the initial stages of the expansion and the final stages of the contraction the metric is singular  $\tilde{a}(t)  \propto~  t^{1/2} \ln^{1/4}{1\over t^{1/2}}$. Up to the logarithmic  term the singularity is similar to that in the cosmological models with relativistic matter where $ \tilde{a}(t) \propto  ~  t^{1/2}$.  The difference is that here the scale factor is periodic in time while in open cosmological model ($k=-1$) with relativistic matter the expansion is eternal.  Thus when $a^2_0 >  L^2 / \gamma^2_c $  (\ref{TypeIcond}),  the vacuum energy density $\epsilon$ is able to reverse the initial expansion shown in Fig.\ref{figU-1}.

The field strength (\ref{fieldstrength}) evolution in time is expressible in terms of $b(\tau)$  function:
\be\label{fieldminI}
2 g^2 \CF = { e^{b^2(\tau)}  \over \mu^4_1 }   \Lambda^4_{YM}. 
\ee
The minimum  value of the field strength   (\ref{fieldminI}) at $\tau =\tau_m$ ($b =0$) is  
\be
2 g^2 \CF_{m} =  { 1 \over \mu^4_1 }  \Lambda^4_{YM}, ~~~~
 \ee
and from (\ref{solk-1-I}), (\ref{limsI}) it follows that the minimum value  of the field strength  varies in the interval
\be
 { 1 \over e }  \Lambda^4_{YM}  < 2 g^2 \CF_{m} \leq e \Lambda^4_{YM}
\ee
when $\gamma^2 \in [0, \gamma^2_c)$.
The energy density and pressure  (\ref{energypressure3}) will evolve in time as well:
\beqa
\epsilon =       {  \CA  \over \mu^4_1 } e^{b^2(\tau)}  \Big( b^2(\tau)- \gamma^2 \mu^2_1 \Big) \Lambda^4_{YM} ,~~~p =    {  \CA  \over 3 \mu^4_1 } e^{b^2(\tau)}  \Big(  b^2(\tau)- \gamma^2 \mu^2_1 +4 \Big) \Lambda^4_{YM}.  
 \eeqa
The equation of state will take the following form: 
 \be
 p = {\epsilon \over 3} +  ~{ 4 \CA  e^{b^2(\tau)}    \over 3 \mu^4_1 }  \Lambda^4_{YM}~ >~ {\epsilon \over 3},~~~~~~~~~~
 w_I={p\over \epsilon}={b^2(\tau)- \gamma^2 \mu^2_1 +4  \over  3 \Big( b^2(\tau)- \gamma^2 \mu^2_1 \Big)}
 \ee
showing that the pressure is larger than that in the case of radiation and $  w_I  \in (-1,  {1\over 3})$. The values of the energy density and pressure at $\tau = \tau_m$ are
\beqa
 \epsilon_{m} =     - \CA  { \gamma^2 \mu^2_1 \over \mu^4_1   }    \Lambda^4_{YM} ,~~~~~
p_{m} =     \CA  { 4-\gamma^2 \mu^2_1\over 3 \mu^4_1   }    \Lambda^4_{YM} .
\eeqa
\begin{table}[htbp]
   \centering
     \begin{tabular}{@{} cccccc @{}} 
     
      \toprule
      $\gamma^2$ &  $\mu^2_1  $ &  $2 g^2 \CF_{m}/\Lambda^4_{YM}$ & $ \epsilon_{m}/\Lambda^4_{YM}$  & $ p_{m}/\Lambda^4_{YM}$     
      \\
      $\gamma^2 =\Big({L \over a_0}\Big)^2$     &~ $\tilde{a }\in[0,\mu_1]$    &${ 1 \over \mu^4_1 }  $& $ -\CA  { \gamma^2 \mu^2_1\over \mu^4_1   }  $&  $  \CA  { 4-\gamma^2 \mu^2_1 \over 3 \mu^4_1   } $  
      \\  
      \midrule
       0                    &~~${1\over \sqrt{e}}$             & $ e  $         &$0$        &  ${ 4 e\CA  \over 3 }   $       
       \\
   $\gamma^2_c $      &$\sqrt{e}$                &${1\over e}  $       &$- {2  \CA \over e}       $        & $   { 2 \CA  \over 3 e  } $   
  \\
      \bottomrule
   \end{tabular}
   \caption{The limiting values of the variables when $k=-1$ and  $\gamma^2 \in [0, \gamma^2_c)$. The saddle point   value is $\gamma^2_c = 2/ \sqrt{e}$.  In the limit  $\gamma^2 \rightarrow 0$  the   solution (\ref{solk-1}) reduces to the $k=0$  case (\ref{solutk0}), (\ref{bk0}). }
   \label{tbl:largeS1}
\end{table}
 In Table \ref{tbl:largeS1} there are the limiting values of the  field strength, energy density and pressure in the interval  $ \gamma^2 \in [0, \gamma^2_c) $.      The deceleration parameter in Type I case is positive
\be\label{de-accelerationI}
q_I= {b^2+ 2  - \gamma^2 \mu^2_1     \over b^2 - \gamma^2 \mu^2_1 (1 -e^{-b^2/2} )   } \geq 1, 
\ee
where $\gamma^2 \mu^2_1 < 2$   (\ref{limsI})\footnote{ In the formal limits when $\gamma^2 \rightarrow 0$ this expression reduces to the $k=0$ case (\ref{de-acceleration0}) and when $\gamma^2 \rightarrow -\gamma^2$ it reduces to the $k=1$ case (\ref{de-acceleration1}).}.
The Hubble parameter  and the density parameter  $\Omega$ (\ref{omega}), (\ref{omegavac}) are:
\be
L^2 H^2  = {  e^{b^2} \over  \mu^4_1} \Big( b^2 - \gamma^2 \mu^2_1  (1 -e^{-b^2/2} ) \Big)  ,~~~~\Omega_I -1 =  - {\gamma^2 \mu^2_1 e^{-b^2/2}  \over b^2 -  \gamma^2 \mu^2_1 (1 -e^{-b^2/2} ) }. 
\ee
At the typical value of $\gamma^2 =1$  $ \mu^2_1 =  -2 W_0(-{1\over 2\sqrt{e} }) =1$  and  $\tau_m \simeq 0.83$.  With the help of the formulas (\ref{conformaltime}) and (\ref{basicpar}) one can get a numerical estimate  of the expansion proper time interval: 
\be\label{km1per}
 c t_m= \tau_m L     \simeq    1.03 \times 10^{25 }  \Big({eV \over \Lambda_{YM} }\Big)^2 ~ cm, 
\ee
 where $L = 1.25 \times 10^{25} \Big({eV \over  \Lambda_{YM} }\Big)^2 cm$. In comparison,  the Hubble length  is $c H^{-1}_{0} \simeq 1.37\times 10^{28} cm  $. The physical meaning of this result is that the Yang-Mills vacuum energy density $\epsilon$ is able to reverse the  expansion earlier than the Hubble time. In order to be consistent with  the cosmological observational data when  Type I solution is analysed one should have a  typical energy scale of the dimensional transmutation in Yang-Mills theory $\Lambda_{YM}$ to be constrained by  a few electronvolt:  $\Lambda_{YM}  \sim eV$.  This scenario can be realised if the Callan-Symanzik beta-function coefficient $\CA$ in (\ref{mattergroupparameter}),  is effectively small,  like in conformal gauge field theories, and therefore the  scale $L$ is  large enough:
 \be
 L^2 = { 3 c^4 \over 8 \pi G }\  {1\over  \CA\  \Lambda^4_{YM}} \rightarrow \infty,~~~~~\CA \rightarrow 0.
 \ee
 Considering $ \Lambda_{YM} $ in (\ref{basicpar}) to be one of the fundamental interaction scales we obtain the following  lengths: 
 \be\label{scales}
 L_{QCD} \sim10^{9} cm,~~~~~ L_{EW} \sim10^{3} cm,~~~~~L_{GUT} \sim10^{-25}cm,~~~~~L_{Pl} \sim10^{-31} cm.
 \ee
These scales lead to a much shorter universe live-time,  but, importantly, to finite non-diverging  time intervals \footnote{A simplified  direct calculation of the diverging zero-point energy density discussed in the introduction leads to the instant collapse of the universe \cite{Donoghue:2020hoh}.}.  In the case of Type II solution to be considered in the next section, when $0 \leq \gamma^2  < \gamma^2_c $ and $ \tilde{a}(0) \geq \mu_2$, the deceleration parameter $q$ is negative and the scale factor grows exponentially with the inflation of a finite duration (\ref{inflation} ) that undergoes a continuous transition to a linear in time growth (\ref{linear}).

\section{\it Type II Solution }

For the Type $II$ solution we have $0 < \gamma^2  < {2\over \sqrt{e}}$ and $ \tilde{a} \geq   \mu_2 $. The Lamber-Euler  
equation (\ref{Lamber-Euler})
\be
~~~~u e^{-u} = {\gamma^2 \over 2 \sqrt{e}} \nn
\ee
has an alternative solution expressible in terms of $W_-(x)$ function, which represents the other branch of the general $W(x)$ function of the real argument $x$ (see Appendix A).   For the negative values of the argument in the interval $ -1/e \leq x \leq 0$  the function acquires negative values in the interval $ -\infty  \leq W_-(x)  \leq -1$.  Thus the solution takes the following form:
\be
u = -W_{-}\Big(-{\gamma^2 \over 2 \sqrt{e}} \Big).\nn
\ee
The minimal value of the scale factor (\ref{mu-u-eq}) therefore is
\be\label{solak-1-II}
\mu^2_2 = -{2\over \gamma^2} ~W_{-}\Big(-{\gamma^2 \over 2 \sqrt{e}}\Big),~~ ~~~~
\ee
and  it follows that  (see Appendix A)
\be\label{limsII} 
\sqrt{e} < \mu^{2}_{2}  \leq \infty, ~~~~~~~~~~2 < \gamma^2 \mu^2_2 . 
\ee
The interval in which $ \tilde{a}$ takes its values is now infinite:   
\be\label{interval-II}
~~~\tilde{a} \in [\mu_2, \infty].
\ee
With the substitution  
\be\label{scalefactor}
   \tilde{a}^4  =\mu^4_2 e^{b^2},~~~~  ~ b \in  [0,\infty],
\ee
the equation (\ref{k2case})  will take the following form:  
\be\label{bfactor}
{d b \over d \tau}  =   {2 \over \mu^2_2}  ~   e^{-{b^2 \over 2}}   \Big( {\gamma^2  \mu^2_2 \over b^2}  ( e^{{b^2 \over 2}} -1) -1  \Big)^{1/2}.
\ee
With the boundary conditions at $\tau =0$ where $b(0)=0$  ($\tilde{a}(0) = \mu_2$) 
we will get the integral representation of the function $b(\tau)$:
\beqa\label{solk-1-II}
&&\int^{b(\tau)}_{0}    { d b ~  e^{{b^2 \over 2}}   \over 
  \Big( { \gamma^2  \mu^2_2 \over b^2}  (e^{{b^2 \over 2}} -1)- 1 \Big)^{1/2}        } ~ = {2 \over \mu^2_2}   ~ \tau.
\eeqa
The time interval is $\tau \in [0,\infty]$, and as $\tau \rightarrow \infty$,  we have 
\be
b^2(\tau) \simeq 4 \ln{\gamma \over \mu_2}\tau , ~~~~~~a = a_0 \tilde{a} \simeq a_0 \gamma \tau =ct .
\ee
The field strength evolution in time is expressible in terms of $b(\tau)$  function:
\be\label{solk-1-III}
2 g^2 \CF = { e^{-b^2(\tau)}  \over \mu^4_2 }   \Lambda^4_{YM}.
\ee
\begin{figure}
 \centering
\includegraphics[angle=0,width=5cm]{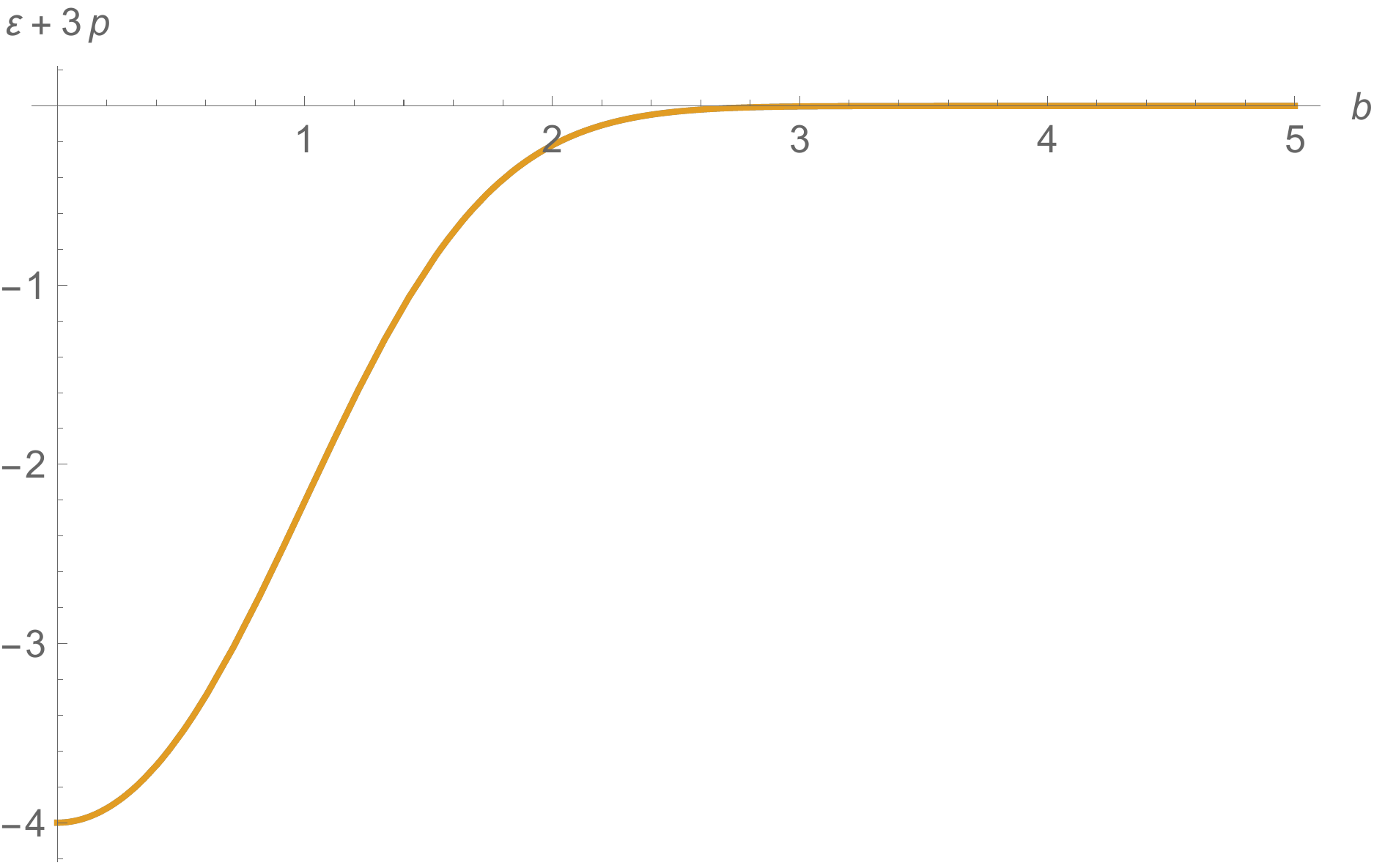}~~~~~
\centering
\caption{   The  r.h.s $\epsilon  +3p$ of the Friedmann acceleration equation 
(\ref{accel})  always  negative in the case of Type II solution (\ref{strongenergycondII}).
}
\label{accelerationII}
\end{figure}
The maximal  value of the field strength (\ref{fieldstrength}) is at $\tau =0$ where  $b(0)=0$:  
\be
2 g^2 \CF_{m} =  { 1 \over \mu^4_1 }  \Lambda^4_{YM}, ~~~~
 \ee
and from (\ref{limsII} )
\be
0 \leq 2 g^2 \CF_{m}  < { 1 \over e }  \Lambda^4_{YM}.  
\ee
 The behaviour of the energy density and pressure is:
\beqa
\epsilon =       - {  \CA  \over \mu^4_2 } e^{-b^2(\tau)}  \Big( b^2(\tau)+ \gamma^2 \mu^2_2 \Big) \Lambda^4_{YM} ,~~~p =   - {  \CA  \over 3 \mu^4_2 } e^{-b^2(\tau)}  \Big( b^2(\tau)+ \gamma^2 \mu^2_2 - 4 \Big) \Lambda^4_{YM},
 \eeqa
and  as  $\tau \rightarrow \infty$  the energy density and pressure tend to zero values of the perturbative vacuum state.  The right-hand side of the Friedmann acceleration equation (\ref{accel}) has  the following form:
\be\label{strongenergycondII}
  \epsilon  +3p =  -  { 2\CA \over \mu^{4}_{2}  } ~ e^{-b^2(\tau)} (   b^2(\tau)+\gamma^2 \mu^2_2 -2  )  \Lambda^4_{YM}, ~~~~~~ b \in  [0,+\infty],
\ee
and is always negative Fig.\ref{accelerationII}. At the initial stages of the expansion  $\tau =0$ ($b=0$) the energy density and pressure are finite and the solution avoids a singular behaviour   $$a(0) = a_0~ \tilde{a}(0) = a_0 ~\mu_2 ~e^{b(0)^2/4}= L {\mu_2\over \gamma} ~> ~0.$$ This behaviour of the scale factor can be compared with the nonsingular solution discussed in \cite{Starobinsky:1980te}. For the  equation of state $p=w \epsilon$ one can find the behaviour of the effective parameter $w$
 \be\label{effectivew}
 w_{II}= {   b^2(\tau)+ \gamma^2 \mu^2_2 - 4  \over  3 \Big( b^2(\tau)+ \gamma^2 \mu^2_2 \Big)  },~~~~~~~-1  \leq  w_{II},
\ee 
where $b \in  [0,\infty]$.  The deceleration parameter of the Type II solution is always negative:
\be\label{de-accelerationII}
q_{II}= 
{b^2 + \gamma^2 \mu^2_2   -2   \over b^2+  \gamma^2 \mu^2_2   (1-e^{b^2/2} )   }
  < 0
\ee
in the  region II (\ref{limsII}) where $2 < \gamma^2 \mu^2_2$.    
 As it follows from (\ref{de-accelerationII}) and  (\ref{solk-1-II}), there is a period of strong  acceleration 
 \be 
 q_{II} \propto -{2\over b^2}
 \ee
 at the initial stages of the expansion $b^2 \sim \tau$  and the scale factor (\ref{scalefactor})  grows exponentially:  
\be\label{inflation}
~~~~~~~a(t)  
\simeq   L ~{\mu_2 \over \gamma}~\exp{\Big[   {2 \over \mu^2_2}  ~ \sqrt{{  \gamma^2 \mu^2_2  \over 2} -1 }  ~ {c t \over L} \Big]}.
\ee  
The inflation is slowing down when  $ct > L $ because  $b^2 $ increases and the acceleration  drops:
\be\label{duration}
q_{II} \propto -{b^2 \over \gamma^2 \mu^2_2} e^{-b^2/2} \rightarrow 0.
\ee
The regime of the exponential growth will continuously  transformed into the   linear in time growth of the scale factor\footnote{The asymptotic solution of  (\ref{bfactor}) is ${b^2 \over 4} \simeq \ln{\gamma \over \mu_2} \tau$ and $a =a_0 \mu_2 \exp{(b^2/4)}$,  as it follows from  (\ref{dimless}), (\ref{scalefactor}).}
\be\label{linear}
a(t) \simeq    ~ c t, ~~~~~~~a(\eta)  \simeq a_0 e^{\eta} .
\ee
The acceleration has its trace on the behaviour of Hubble parameter, which has the following form:
\be
 L^2  H^2 = {  e^{-b^2} \over\mu^4_2} \Big( \gamma^2 \mu^2_2  (e^{b^2/2} -1) -b^2 \Big)  . 
\ee
The $L^2  H^2$ is sharply increasing from zero value and reaches its maximum at 
\be
b^2_s = 1-\gamma^2 \mu^2_2 - 2 W_{-1}\Big(-{\gamma^2 \mu^2_2 \over 4} \exp{({1-\gamma^2 \mu^2_2\over 2})}\Big)
\ee
and allows to estimate its duration    
\beqa\label{stop}
&&  \tau_s =  {  \mu^2_2 \over 2}  \int^{9 b_s}_{0}    { d b ~  e^{{b^2 \over 2}}   \over 
  \Big( { \gamma^2  \mu^2_2 \over b^2}  (e^{{b^2 \over 2}} -1)- 1 \Big)^{1/2}        } ~ .
\eeqa
The number of e-foldings for the  time evolution from $\tau=0$ to $\tau_s$ is defined as
$
\CN = \ln{a(\tau_s) \over a(0)  }  .
$
For the typical parameters  around $\gamma^2 =1.211$, $\mu^2_2 \simeq  1.75$  we get $\tau_s = 10^{23}$ and $\CN  \simeq 53$.
The duration of the inflation  in the case of the GUT scale $\Lambda_{YM}= \Lambda_{GUT}= 10^{16} GeV$ is of  order
\be\label{inflationchartime}
t^{GUT}_s = { L_{GUT} \over c} \tau_s \simeq 4.2 \times 10^{-13 }   ~ sec ,
\ee 
where $L_{GUT} \simeq 1.25 \times 10^{-25} cm $ as in (\ref{scales}). The initial and finale values of the scale factor are: 
$$a(0) =L_{GUT} {\mu_2 \over \gamma} \simeq 1.5 \times 10^{-25} cm,~~~~~~ a(t_s) =L_{GUT} {\mu_2 \over \gamma} e^{\CN}\simeq 1.25 \times 10^{-2} cm, $$
 where $a(t_s)$ is "about the size of a marble" \cite{Guth:1980zm}. 
 The density parameter $\Omega$ (\ref{omega}) has the following form 
\be
~\Omega_{vac} -1 =  -{\gamma^2 \over ({d \tilde{a} \over d \tau})^2 } = - {\gamma^2 \mu^2_2 e^{b^2/2} \over  \gamma^2 \mu^2_2 (e^{b^2/2} -1)  - b^2 }
\ee
and at  $t \gg  t_s$~ ($b^2 \rightarrow \infty$) the vacuum density tends to zero $\Omega_{vac} \rightarrow 0$ meaning that the influence of the gauge field theory vacuum on the evolution of the universe fades out turning into a linear expansion (\ref{linear}).  

It seems natural to include the energy densities $\epsilon_f$  that can contribute into the total energy density $\epsilon =\sum \epsilon_f$ from the hierarchy of fundamental interaction scales.  Taking into account the fact that at each scale (\ref{scales}) the acceleration has a finite duration (\ref{duration}) and appears at a different epoch of the universe expansion, its seems possible that a very large  scale $\Lambda^{'}_{YM} \gg GeV$ contributes to the inflation at the initial stages of the expansion and a smaller scale $\Lambda^{''}_{YM} \simeq eV$ contributes to the late-time acceleration of the universe.  In addition here we do not include the energy density of the standard matter (\ref{standradmatter}) that can be easily included, and the subsequent evolution of the universe will turn into the standard hot universe expansion.  In the next section we will consider the Type III solution when the parameter $\gamma^2$ is equal to its critical value  $\gamma^2  = \gamma^2_c$.

\section{\it Type III Solution (Separatrix) }

Consider now the Type  $III$ solution when $ \gamma^2  = \gamma^2_c = {2\over \sqrt{e}}$. The Lamber-Euler  
equation (\ref{Lamber-Euler})
\be
~~~~u e^{-u} = {\gamma^2_c \over 2 \sqrt{e}} = {1  \over e} \nn
\ee
in this case has a unique solution   (see Appendix B)
\be
u_c =-W_0(-{1\over e})= -W_{-}(-{1\over e})=1,\nn
\ee
and from (\ref{mu-u-eq}) we get  
\be\label{mus}
 \mu^2_c = {2 u_c \over \gamma^2_c}   =  \sqrt{e} ~, ~~~~~~~\mu^2_c   \gamma^2_c =2.
\ee
The interval in which  $ \tilde{a}$ variates is now
\be
~~~\tilde{a} \in [0, \mu_c].
\ee
When $\gamma^2 \rightarrow \gamma^2_c  $, the region I and region II (\ref{solk-1-I}) and (\ref{solak-1-II}) merge at $\mu^2_1=\mu^2_2 \rightarrow \mu^2_c$,  as one can see from (\ref{limsI}), (\ref{limsII}).  With the substitution  
\be
   \tilde{a}  =\mu_c  e^{b},~~~~  ~ b \in  [-\infty,0],
\ee
the equation (\ref{k2case})  will take the following form:  
\be
{d b \over d \tau}  =  \sqrt{ {2 \over e} } ~   e^{-2 b}   \Big(   e^{2 b} -1-2b  \Big)^{1/2}.
\ee
With the boundary conditions $b(0)=-\infty$ ($\tilde{a}(0) = 0$) in place  
 we will get the integral representation of the function $b(\tau)$:
\beqa\label{solk-1-III}
&&\int^{b(\tau)}_{-\infty}    { d b ~  e^{2b}   \over 
  \Big(  e^{2b} - 1-2b \Big)^{1/2}        } ~ = \sqrt{{2 \over e}}   ~ \tau,~~~~~~\tau \in [0,\infty].
\eeqa
The field strength evolution in time takes the following form:
\be\label{filstreIII}
2 g^2 \CF = e^{-4 b(\tau)-1}    \Lambda^4_{YM}.
\ee
The behaviour of the energy density and pressure is:
\beqa\label{enegmomenIII}
&&\epsilon   =    2\CA   e^{-4b(\tau) -1}  \Big(-2b(\tau) -1 \Big)\Lambda^4_{YM} , ~~~~~~
 p =    { 2 \CA  \over 3 } e^{-4 b(\tau)-1}  \Big( -2b(\tau) +1 \Big) \Lambda^4_{YM}.
 \eeqa
 \begin{figure}
 \centering
\includegraphics[angle=0,width=5cm]{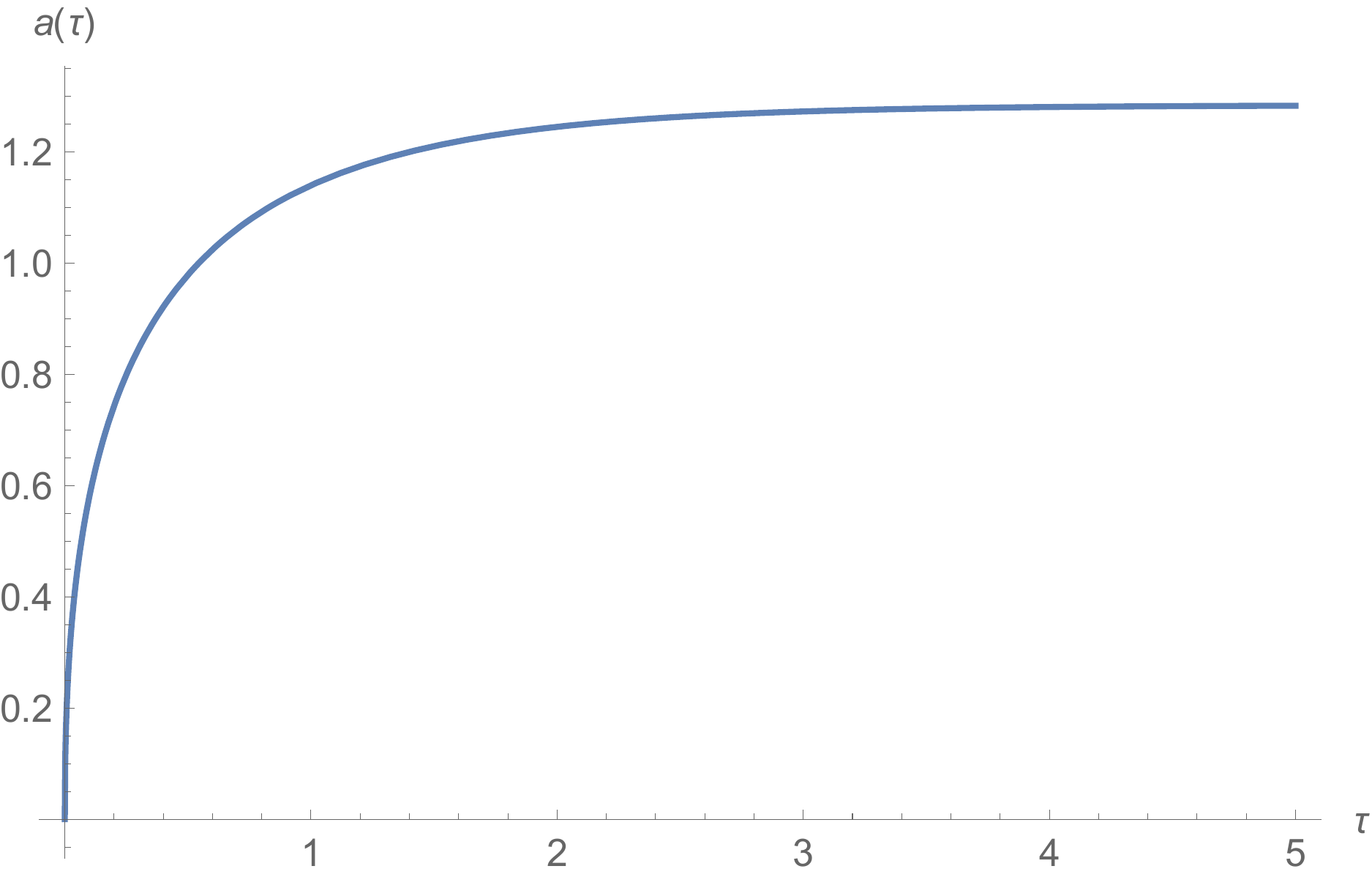}~~~~~
\centering
\caption{When $k=-1$ and $ \gamma^2   =\gamma^2_c= {2\over \sqrt{e}}$, the Type III solution is approaching asymptotically the maximum value $\tilde{a}= \mu_c$ as $\tau \rightarrow \infty$.    
}
\label{saddlepoint}
\end{figure}
There is a characteristic time $\tau = \tau_0 $, corresponding to $b=-1/2$   
 \be
 \int^{-1/2}_{-\infty}    { d b ~  e^{2b}   \over 
  \Big(  e^{2b} - 1-2b \Big)^{1/2}        } ~ = \sqrt{{2 \over e}}   ~ \tau_0    
 \ee
when  the energy density approaches the zero value 
\be 
2g^2 \CF_0 = e \Lambda^4_{YM},~~~~~\epsilon_0 =0,~~~~~p_0 =  { 4 \CA  \over 3 e } \Lambda^4_{YM}.
\ee 
The scale factor asymptotically approaches a maximal static value shown in Fig.\ref{saddlepoint}
\be\label{stationary1}
\tilde{a} = \mu_c  e^{b} \rightarrow  \mu_c  
 \ee
when $\tau \rightarrow  \infty $ and $ b \propto e^{-\sqrt{2} \tau } \rightarrow  0 $ in (\ref{solk-1-III}). 
The energy density  becomes negative in the region $b\in (-1/2,0]$.  The field strength, energy density, and pressure are approaching asymptotically  the following values:  
\be\label{stationary2} 
2g^2 \CF_c = {1\over e}  \Lambda^4_{YM},~~~~~\epsilon_c =- {  2\CA  \over e } \Lambda^4_{YM},~~~~~p_c =  { 2 \CA  \over 3 e } \Lambda^4_{YM}.
\ee
 According to the Friedmann equations (\ref{edens})-(\ref{accel}) the acceleration  is driven by the overall sign of the $\epsilon +3 p$  that can be calculated by using the expressions (\ref{enegmomenIII})
\beqa\label{FriedmannEquationsIII}
\epsilon  +3p =  -  8\CA    b(\tau) e^{-4b(\tau)-1 }   \Lambda^4_{YM} \geq 0,~~~~~~~~~~~b \in  [-\infty,0].
\eeqa 
The strong energy dominance condition $\epsilon +3 p \geq 0$ holds here.  The deceleration parameter for the Type III solution is always positive, $b\in [-\infty, 0]$:
\be\label{de-accelerationIII}
q_{III}=
{b  \over b + {1 \over 2} (1 - e^{2b} )   }  \geq  0.
\ee
The Hubble parameter  and the density parameter $\Omega$ (\ref{omega}) are:
\be
L^2 H^2 = 2 e^{-4 b-1}   \Big( e^{2 b} -1-2b \Big)  ,~~~~\Omega_{vac} -1 =  -{\gamma^2_c \over ({d \tilde{a} \over d \tau})^2 } =  -{  e^{2 b } \over    e^{2 b} -1-2b   }. 
\ee
 The Type III "static" solution is a separatrix.   It is tuned to the critical value $\gamma^2 = \gamma^2_c$ and the infinitesimal  deviation from the critical value turns the solution either into the Type II solution or into the Type IV solution that we will consider in the next section. The Type IV solution, in the parameter region $\gamma^2 > \gamma^2_c $, is characterised by the appearance of a late-time acceleration.

\section{\it Type IV Solution}

\begin{figure}
 \centering
\includegraphics[angle=0,width=7cm]{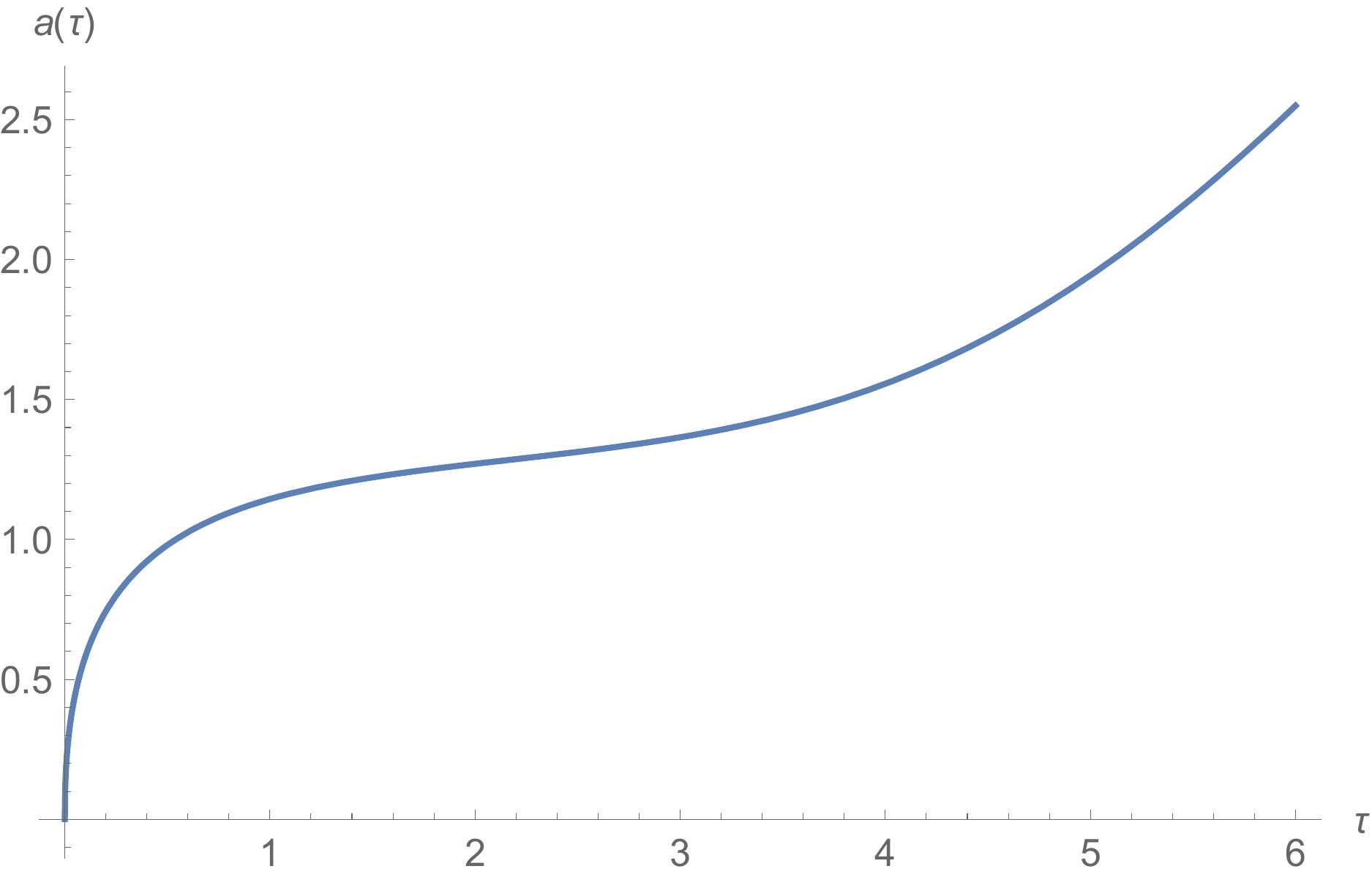}~~~~~
\centering
\caption{At $k=-1$ and $  {2\over \sqrt{e}} <  \gamma^2  $ the value of $\tilde{a}$ is in the interval  $\tilde{a} \in [0,\infty]$. The solution shows four stages of alternating  expansion. In the first stage there is a period of deceleration, in the second stage the expansion reaches a quasi-stationary evolution near $\tilde{a} \simeq \mu_c$,  in the third stage there is a period of exponential expansion of a finite duration that undergoes a continuous transition to the fourth stage of a   linear in time  growth.     
}
\label{abovesaddlepoint}
\end{figure}

\begin{figure}
 \centering
\includegraphics[angle=0,width=7cm]{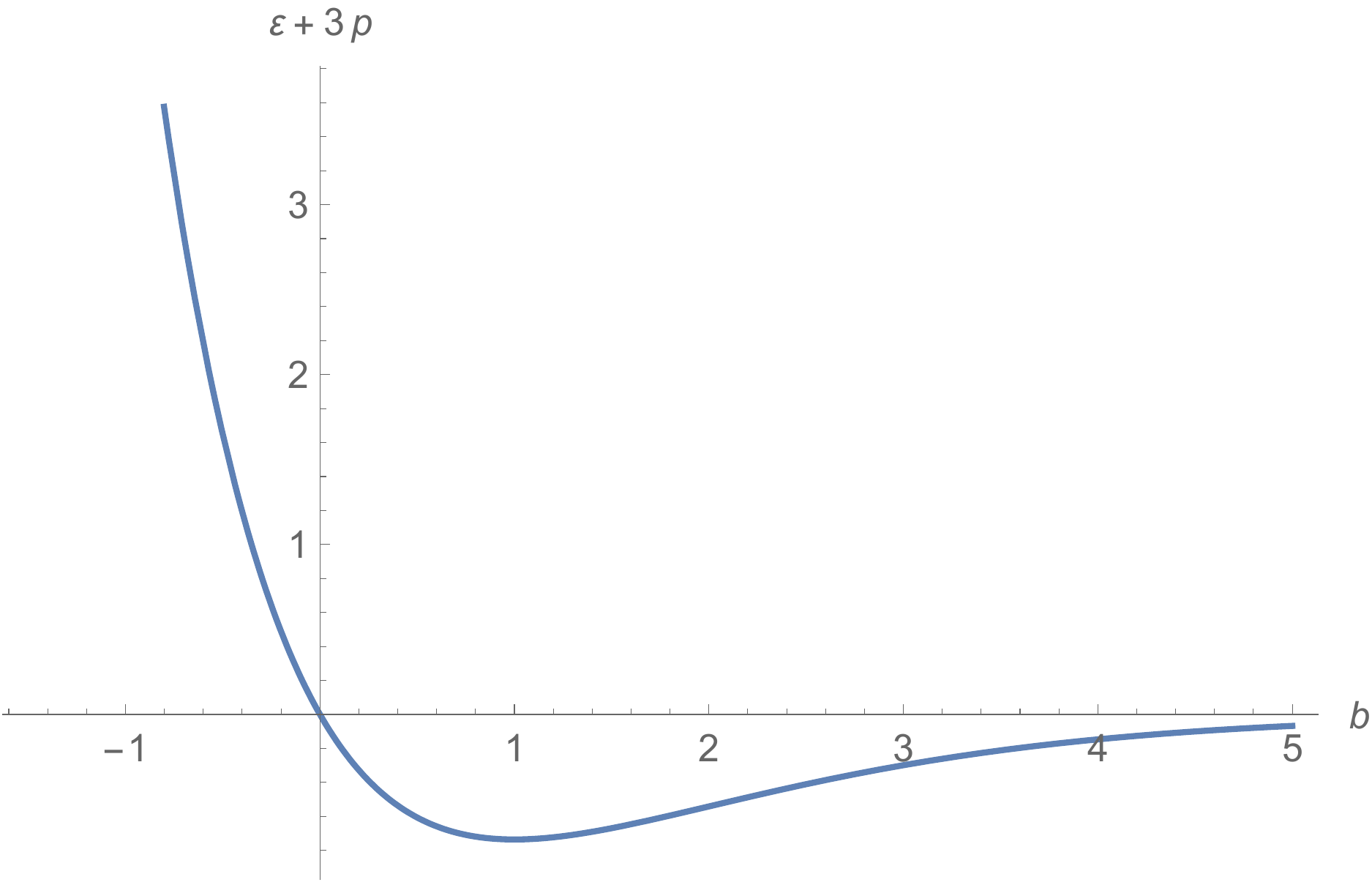}~~~~~
\centering
\caption{   The  r.h.s $\epsilon  +3p$ of the Friedmann acceleration equation 
(\ref{accel})   is positive when $b<0$ and is negative when $b >0$ in the case of Type IV solution (\ref{strongenergycondIV}).
}
\label{acceleration}
\end{figure}

The Type $IV$ solution is defined in the region $\gamma^2 > \gamma^2_c $ where the equation  
\be
U_{-1}(\mu)=  {1 \over \mu^2}  
 \Big( \log{ {1 \over \mu^4 } } -1\Big) + \gamma^2=0~~~~
  \ee
has no real solutions.  The potential function $U_{-1}(\tilde{a})$ is always positive for all positive values of $\tilde{a}$ and the scale factor variates in the whole interval  $\tilde{a} \in [0, \infty]$ (see Fig.\ref{abovesaddlepoint}).  With the substitution  
\be
   \tilde{a}  =\mu_c  e^{b},~~~~  ~ b \in  [-\infty,\infty], ~~~~~~~2 < \gamma^2 \mu^2_c,
\ee
where  $ \mu^2_c =   \sqrt{e} $, as in (\ref{mus}), the equation (\ref{k2case})  will take the following form:  
\be
{d b \over d \tau}  =  \sqrt{ {2 \over e} } ~   e^{-2 b}   \Big(  {\gamma^2 \over \gamma^2_c} e^{2 b} -1-2b  \Big)^{1/2}.
\ee
With the boundary conditions $b(0)=-\infty$  ($\tilde{a}(0) = 0$)    
we will get the integral representation of the function $b(\tau)$:
\beqa\label{solk-1-IV}
&&\int^{b(\tau)}_{-\infty}    { d b ~  e^{2b}   \over 
  \Big(  {\gamma^2 \over \gamma^2_c} e^{2b} - 1-2b \Big)^{1/2}        } ~ = \sqrt{{2 \over e}}   ~ \tau,~~~~~~\tau \in [0,\infty].
\eeqa
The field strength evolution in time is similar to the Type III solution (\ref{filstreIII}):
\be
2 g^2 \CF =   e^{-4 b(\tau)-1}     \Lambda^4_{YM},
\ee
but  the time dependence of $b(\tau)$ is different and is defined now by the  equation (\ref{solk-1-IV}). 
The same is true for the behaviour of the energy density and pressure:
\beqa\label{energydenIV}
&&\epsilon  =   2\CA    e^{-4b(\tau)-1 }  \Big(-2b(\tau) -1 \Big)\Lambda^4_{YM} , ~~~~~~
 p =    { 2 \CA  \over 3  } e^{-4 b(\tau)-1}  \Big( -2b(\tau) +1 \Big) \Lambda^4_{YM}.
 \eeqa
The right-hand side of the Friedmann acceleration equation (\ref{accel}) has a similar expression with the Type III solution (\ref{FriedmannEquationsIII}):
\be\label{strongenergycondIV}
  \epsilon  +3p =  -   8\CA~   b(\tau) e^{-4b(\tau) -1}   \Lambda^4_{YM}, ~~~~~~ b \in  [-\infty,+\infty],
\ee
and the strong energy dominance condition $\epsilon +3 p \geq 0$ is violated here when $b>0$ and 
the region of  positive acceleration is wherefore at  $b > 0$ shown in Fig.\ref{acceleration}. 
 Thus the deceleration parameter for the Type IV solution is sign alternating,  $b\in [-\infty,  \infty]$:
\be\label{de-accelerationIV}
q_{IV}= {b  \over b + {1 \over 2} (1 - {\gamma^2   \over  \gamma^2_c } e^{2b} )   } ,  
\ee
it is positive for $b\in [-\infty,0)$ and is negative for $b\in (0,\infty]$.  Therefore the character of the solution is changing  at $b=0$ where the deceleration parameter  $q_{IV} =0$.  In these two regions the  behaviour of the solution is qualitatively different.  At the quasi-stationary point $\tau = \tau_c$  ($b=0 $)     
 \be
 \int^{0}_{-\infty}    { d b ~  e^{2b}   \over 
  \Big(  {\gamma^2 \over \gamma^2_c}  e^{2b} - 1-2b \Big)^{1/2}        } ~ = \sqrt{{2 \over e}}   ~ \tau_c 
 \ee
we have  
$$ 2g^2 \CF_c = {1\over e}  \Lambda^4_{YM},~~~~~\epsilon_c =- {  2\CA  \over e } \Lambda^4_{YM},~~~~~p_c =  { 2 \CA  \over 3 e } \Lambda^4_{YM},
$$
and it is  reminiscent to the stationary behaviour of the Type III solution (\ref{stationary1}), (\ref{stationary2}).  The energy density (\ref{energydenIV}) is changing its sign at $\tau = \tau_0 $ ($b=-1/2$)
 \be
 \int^{-1/2}_{-\infty}    { d b ~  e^{2b}   \over 
  \Big(  {\gamma^2 \over \gamma^2_c}  e^{2b} - 1-2b \Big)^{1/2}        } ~ = \sqrt{{2 \over e}}   ~ \tau_0    
 \ee
where we have  
$$ 
2g^2 \CF = e \Lambda^4_{YM},~~~~~\epsilon =0,~~~~~p =  { 4 \CA  \over 3 e } \Lambda^4_{YM}.
$$  
Thus there are four stages of alternating  expansions. There is a period of  deceleration in the first stage $\tau \ll \tau_c$ where $q_{IV} $ is positive.  In the second stage, in the vicinity of  $\tau \sim \tau_c$ where $q_{IV}=0$ the expansion is quasi-stationary and a slow varying scale factor is of order $\tilde{a}(\tau) \simeq \mu_c$.  In the third stage $\tau > \tau_c$ there is a period of exponential expansion of a finite duration $b \sim (0,5)$ where $q_{IV}$ is negative. It is of finite duration because when   $b > 0 $ is large, the acceleration tends to zero:
$$
q_{IV} \simeq - {2  \over \gamma^2 \mu^2_c} b e^{-2b} .
$$ 
In the fourth stage  $\tau \gg \tau_c$, where $e^b \simeq  {\gamma \over \gamma_c} \sqrt{2\over e} \tau$,  the  acceleration drops to zero  $q_{IV} \simeq 0$ and the universe undergoes a continuous transition to a  linear in time growth of the scale factor 
\be\label{linear4}
a(t) \simeq   ~ c t~, ~~~~~a(\eta) \simeq   ~ e^{\eta}
\ee   
and the Hubble parameter (\ref{habble}) has the following  behaviour: 
\be
H =  \sqrt{{2 \over e}}{e^{- 2b} \over L }\Big( {\gamma^2 \over \gamma^2_c}e^{2 b} -1-2b \Big)^{1/2}  \simeq {1\over c t}  . 
\ee 
When $\tau \gg \tau_c$    the  $2g^2 \CF \rightarrow 0$ and  the energy density and pressure are approaching the zero values, $\Omega$ (\ref{omega}) tends to zero value as well:
\be
\Omega_{vac}  = 1 -{\gamma^2 \over ({d \tilde{a} \over d \tau})^2 } = 1 - {\gamma^2   e^{2 b } \over     \gamma^2_c \Big( {\gamma^2 \over \gamma^2_c} e^{2 b} -1-2b \Big) } \rightarrow 0. 
\ee 
The influence of the gauge field theory vacuum on the evolution of the universe is fades out at very late-time.
It seems that  the Type IV solution is useful to  explain a late-time acceleration of the universe expansion if one appropriately adjust the parameters $a_0$ and $\gamma$. 

\section{\it Flat Geometry, $k=0$}

\begin{figure}
 \centering
\includegraphics[angle=0,width=5cm]{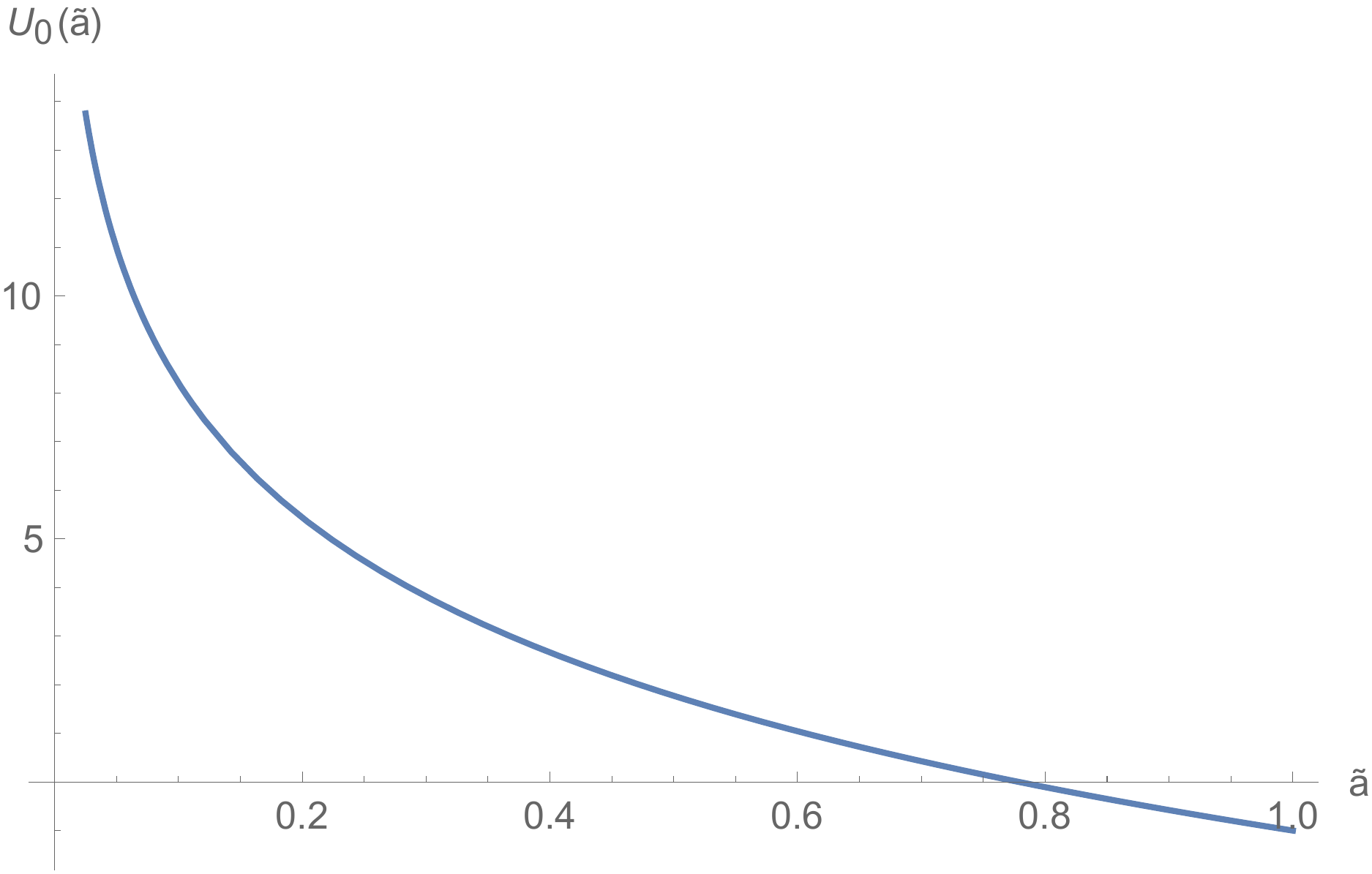}~~~~~
\includegraphics[angle=0,width=5cm]{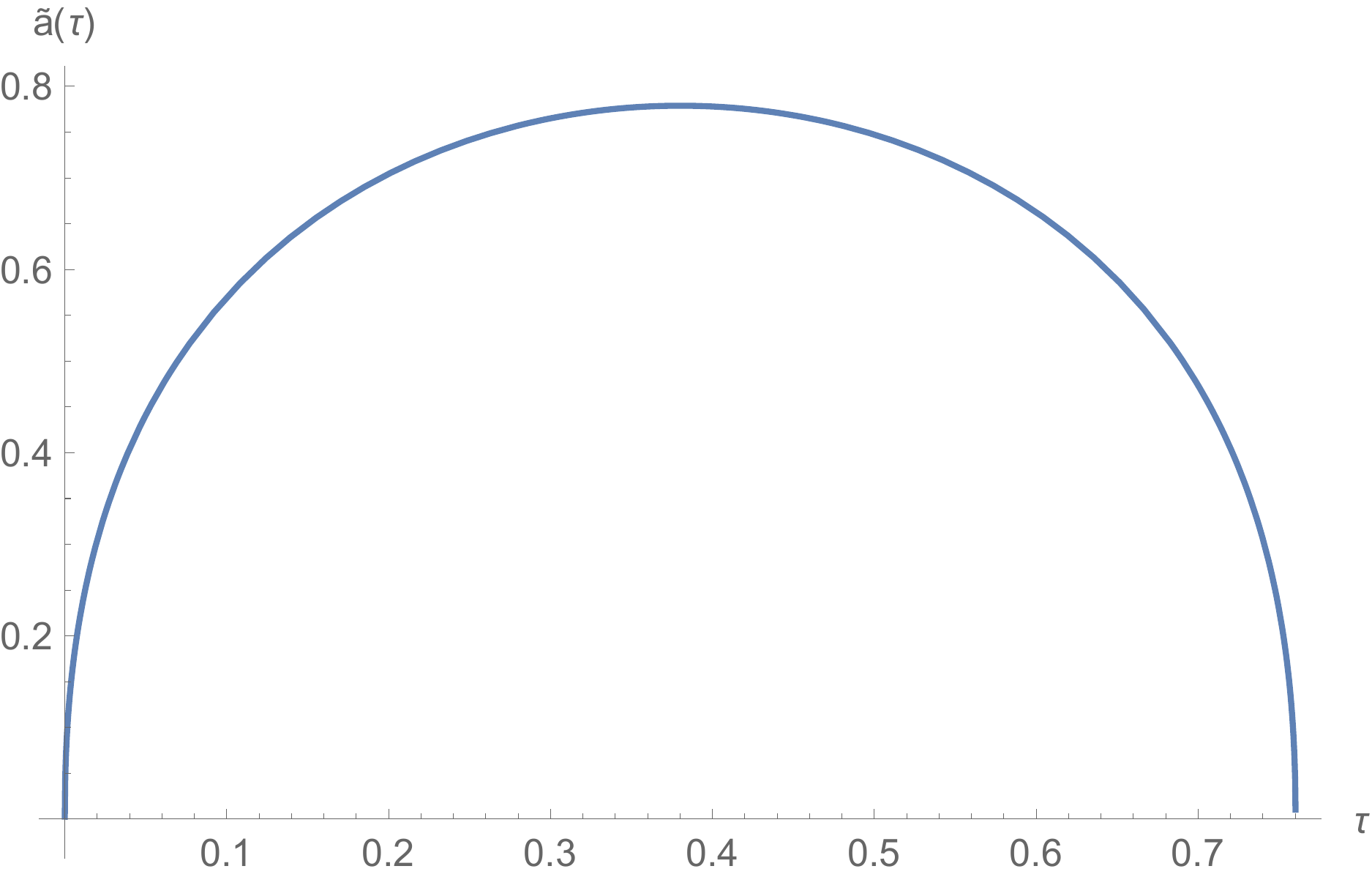}
\includegraphics[angle=0,width=5cm]{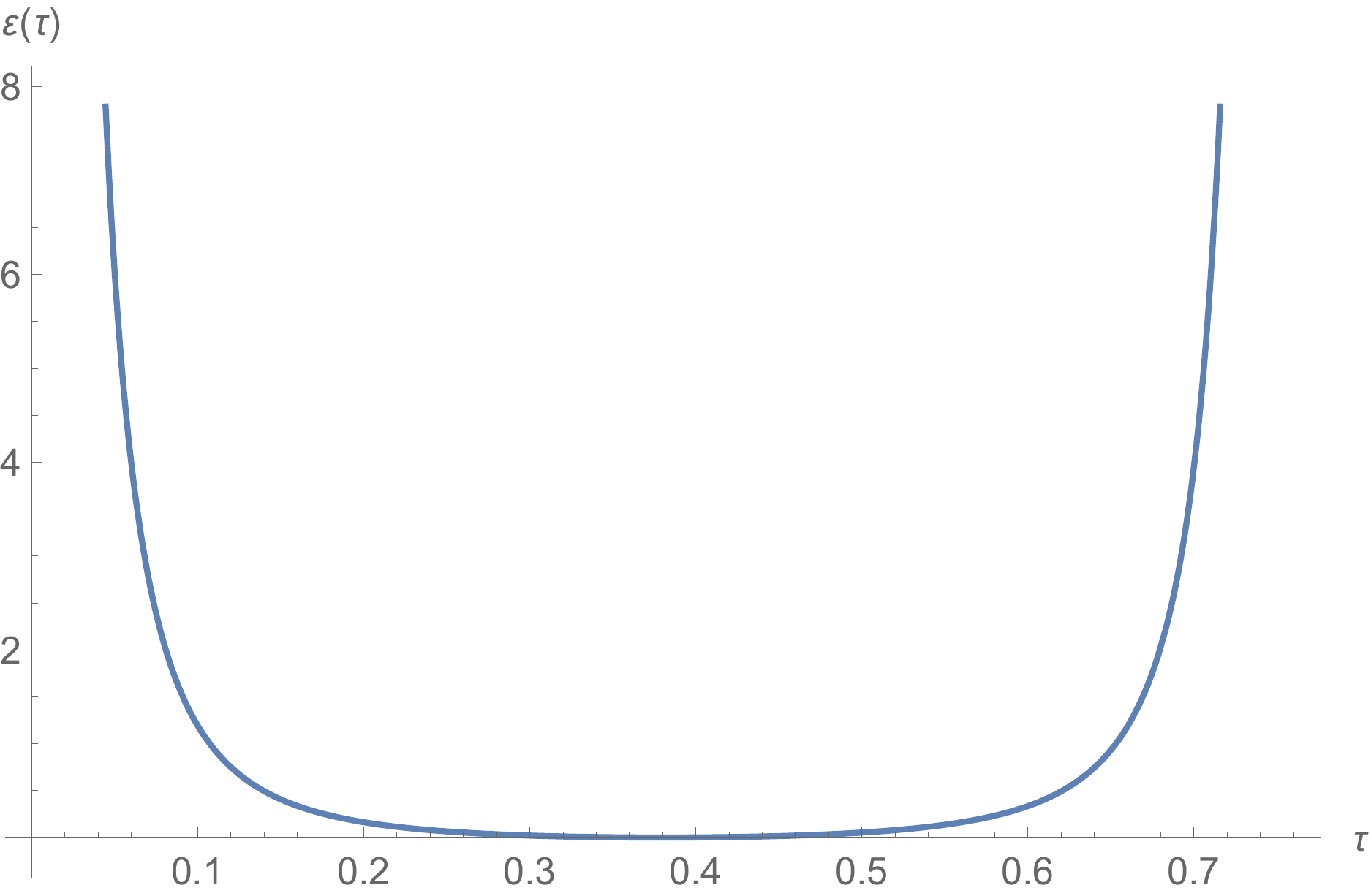}
\centering
\caption{The left figure shows the behaviour of the potential function $U_0(\tilde{a})$. It  is positive for $\tilde{a} \leq \mu ={1\over e^{1/4}} $.   The centre figure shows the behaviour of the scale factor $\tilde{a}(\tau)$. The maximal value of the scale factor at the half period ~$\tau_m =\sqrt{{\pi \over 8 e}} $ is   $\tilde{a}(\tau_{m})=\mu $. The left figure shows the behaviour of the energy density $\epsilon(\tau)$.  }
\label{figU0} 
\end{figure}

The evolution equation (\ref{basiceq}) in this case takes the following form:
\beqa\label{k0case}
{d\tilde{a} \over d \tau }= \pm \sqrt{   {1 \over \tilde{a}^2} \Big(\log{ {1 \over \tilde{a}^4 }} -1\Big)  },~ 
\eeqa
and  the "potential"  function is (see Fig.\ref{figU0})
\be
~~~~U_0(\tilde{a}) \equiv    {1 \over \tilde{a}^2    }  
\Big(  \log{ {1 \over \tilde{a}^4    }} -1 \Big).
\ee
The solution of the equation $U_0(\mu) =0$ determines the values of the scale factor $\tilde{a}=\mu$ at which the square root  changes its sign.  The evolution equation (\ref{k0case})  should be restricted to those  real values of $\tilde{a} \in [0,\mu]$ at which the potential $U_{0}(\tilde{a}) $ is nonnegative. The maximal value of the scale factor $\tilde{a}_{m} \equiv \mu$ is defined by the equation  
\be\label{bound}
 U_{0}(\mu)= {1 \over \mu^2}  
 \Big( \log{ {1 \over \mu^4 } } -1\Big) =0,~~~~\mu^2 = {1\over \sqrt{e}}.
  \ee
With the substitution  
\be\label{atilde}
 \tilde{a}^4   = \mu^4 e^{-b^2}, ~~
\ee
where $b \in [-\infty,\infty]$, the equation (\ref{k0case}) will take the following form: 
\be\label{equationk0}
~~~~{d b \over d \tau} =  {2 \over \mu^2 } ~   e^{{b^2\over 2 }}.
\ee
Integrating the equation (\ref{equationk0}) with the boundary conditions $a(0) = 0$,  $b(0)=-\infty$ we find
\beqa\label{solutk0}
 \int^{b(\tau)}_{-\infty} d b     e^{-{b^2\over 2 } }= {2 \over \mu^2 } ~ \tau,~~~~~~~~\tau \in [0, 2 \tau_m]
\eeqa
where the half period is
\be\label{expantiontime}
\tau_m =  {\mu^2 \over 2} \int^{0}_{-\infty} d b   e^{-{b^2\over 2 } }= \sqrt{{\pi \over 8 e}}.
\ee
The solution can be expressed in terms of the inverse error function $InverseErf(x)$ \cite{Abramowitz}
\be\label{bk0}
b(\tau) = \sqrt{2} InverseErf\Big(\sqrt{ {8 e \over \pi  } }~(\tau-\tau_m) \Big) 
\ee
and for $\tilde{a}$ in (\ref{atilde}) we will get the solution shown in Fig.\ref{figU0}:
\be\label{solutionk0}
\tilde{a}(\tau)= \exp{\Big( -{1\over 4} - {1\over 2}  InverseErf^2\Big(\sqrt{ {8 e \over \pi} }~(\tau-\tau_m) \Big) \Big) }.
\ee
The scale factor is periodic in time (\ref{solutionk0}) as it is in closed Friedmann  cosmological model (k=1). The asymptotic behaviour of  the $\tilde{a}(\tau)$ near $\tau=\tau_m$ and $\tau =0$  has the form 
\beqa\label{singu}
\tilde{a}(\tau) \simeq   \begin{cases}  e^{-1/4} - e^{3/4}  (\tau-\tau_m)^2,~~~~~\tau \rightarrow \tau_m \\ 2  \tau^{1/2} \ln^{1/4}{1\over \tau^{1/2}},~~~~~~~~~~~~~\tau \rightarrow 0   \end{cases},
\eeqa
that is, $\tilde{a}(t)  \propto~  t^{1/2} \ln^{1/4}{1\over t^{1/2}}$. The singularity is logarithmically weaker compared to that in cosmological model with relativistic matter where $ \tilde{a}(t) \propto  ~  t^{1/2}$. For the  field strength we have (\ref{fieldstrength})
\be
2 g^2 \CF =  e^{b^2+1} \Lambda^4_{YM}
\ee
and for the energy density (\ref{atilde}) and pressure  (\ref{energypressure3}) evolution in time is   
\be 
 \epsilon =       \CA  b^2 e^{b^2+1} \Lambda^4_{YM} ,~~~p =     { 4 \CA \over 3  }    e^{b^2+1} \Lambda^4_{YM} ,
\ee
where $b(\tau) $ is given in (\ref{bk0}) and $\tau = c t /L$.  The behaviour of $\epsilon $ is shown in Fig.\ref{figU0}. At the half period $\tau=\tau_m$ where $\tilde{a} = \tilde{a}_{m}$ ($b=0$)  we will have 
\beqa
2 g^2 \CF_{m} =   e \Lambda^4_{YM},~~~~~\epsilon_{m} = 0,~~~~~~~~
p_{m}= {4 \CA \over 3} ~ e \Lambda^4_{YM}.
\eeqa
At the beginning of the expansion from (\ref{singu}) we get
\be
 2 g^2 \CF  \propto  {1 \over t^2 \ln{1\over t^{1/2}}} ,~~~ \epsilon  \propto {1 \over t^2 \ln{1\over t^{1/2}}} \ln\Big[{1 \over t^2 \ln{1\over t^{1/2}}}\Big]  , ~~~~p \propto {1 \over t^2 \ln{1\over t^{1/2}}} \ln\Big[{1 \over t^2 \ln{1\over t^{1/2}}}\Big].
\ee
The deceleration parameter (\ref{deceleration}) here is always positive:
\be\label{de-acceleration0}
q = 1 +{2 \over b^2 } \geq 1, 
\ee
where we used  (\ref{atilde}) and (\ref{bound}).  The Hubble parameter  and the density parameter $\Omega$ (\ref{omega}) are:
\be
H^2 = {\dot{a}^2 \over a^2}= {b^2 \over L^2}  e^{b^2+1},~~~~\Omega_{vac} =  {8\pi G \over 3 c^4} {\epsilon \over H^2} =1. 
\ee
The expansion proper time interval (\ref{conformaltime}) in this case of flat geometry $k=0$ is (\ref{expantiontime}):
\be\label{k0per}
~~~~~c t_m  = \tau_m L =  \sqrt{{\pi \over  8e}} L ~   \simeq      4.7 \times 10^{24}    \Big({eV \over \Lambda_{YM} }\Big)^2 ~ cm, 
\ee
where $L = 1.25 \times 10^{25} \Big({eV \over  \Lambda_{YM} }\Big)^2 cm$.   The physical meaning of this result is that the vacuum energy density $\epsilon$ is able to slow down the expansion earlier than the Hubble time $c H^{-1}_{0} =1.37\times 10^{28} cm  $ even in the case of flat geometry ($k=0$).  Here the scale $\Lambda_{YM} $  is of order of a few electronvolt  $\Lambda_{YM}  \sim eV$, as in the case  of the  Type I solution.

\section{\it  Spherical  Geometry, $k=1$ }

\begin{figure}
 \centering
\includegraphics[angle=0,width=5cm]{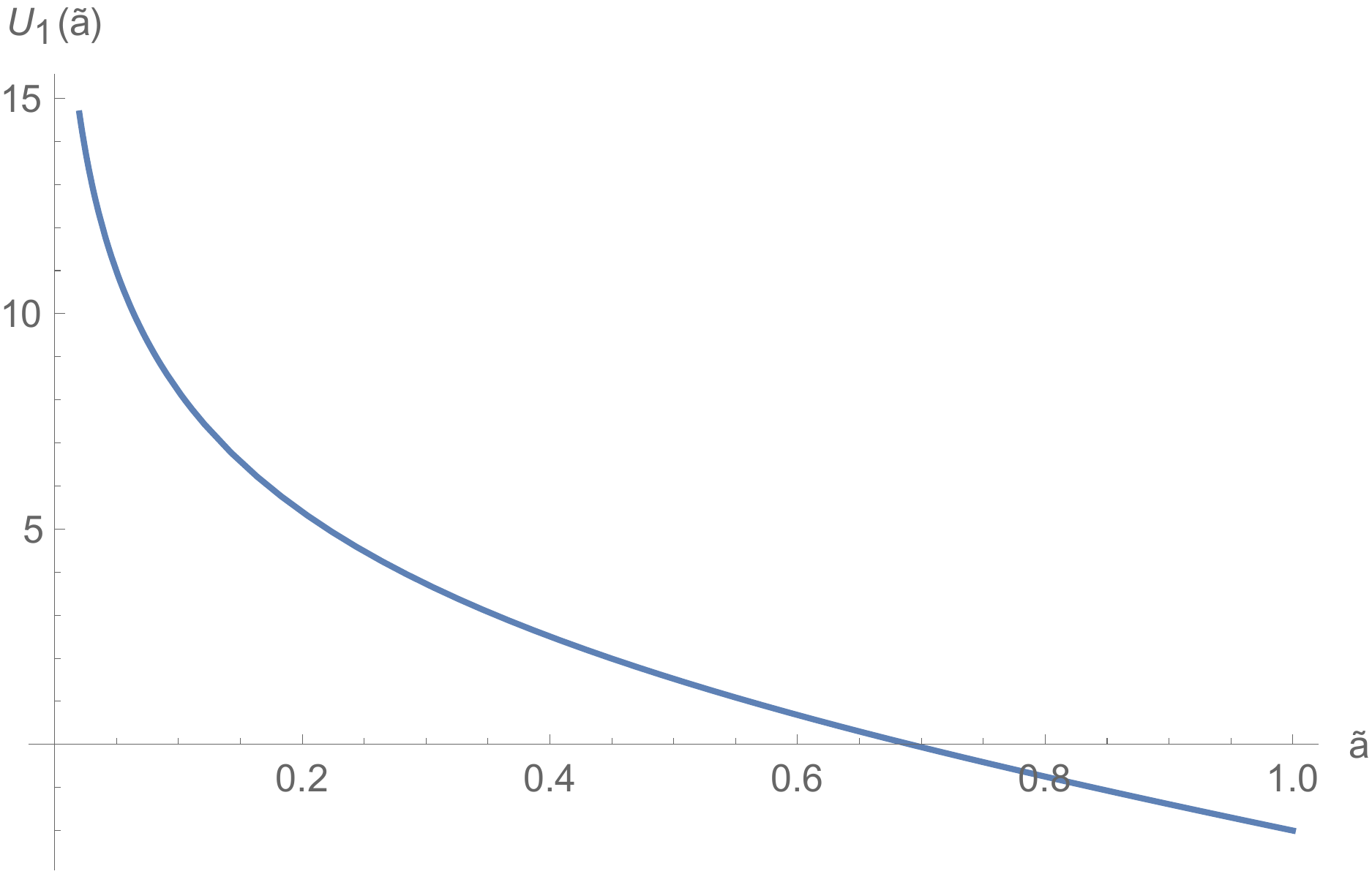}~~~~~
\includegraphics[angle=0,width=5cm]{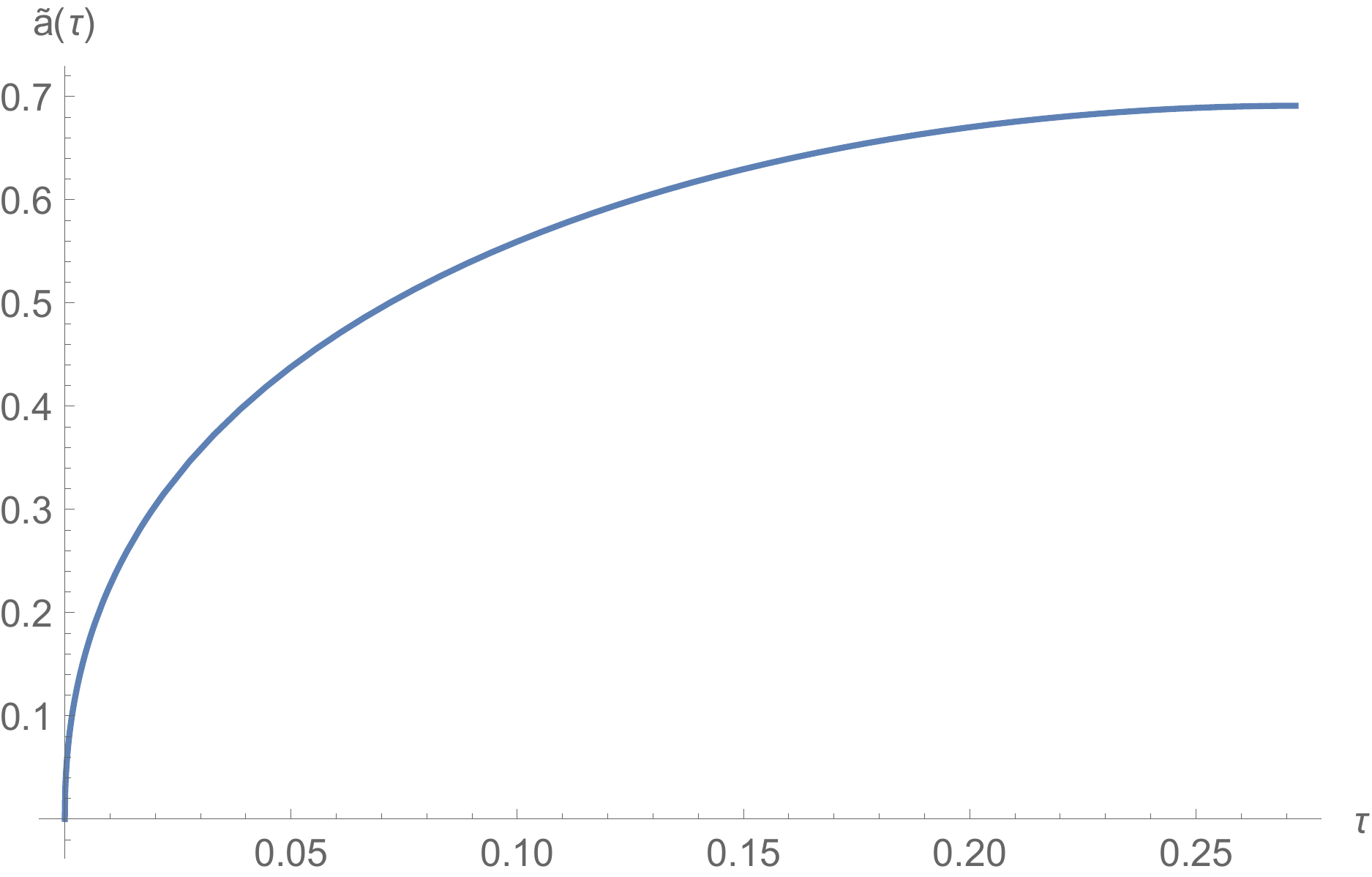}
\centering
\caption{In the case $k=1$ the values of $\tilde{a}$ are in the interval  $\tilde{a} \in [0,\mu]$. When  $\gamma^2 =1$,    $\mu \simeq 0.69$~and ~$\tau_m \simeq   0.27$. }
\label{figU1} 
\end{figure}

The corresponding equation (\ref{basiceq}) will take the following form:
\beqa\label{k1case}
&&{d \tilde{a} \over d \tau}= \pm \sqrt{ {1 \over \tilde{a}^2} \Big( \log{ {1 \over \tilde{a}^4 } } -1\Big) -   \gamma^2 },~~~
\eeqa
and the "potential"  function is
\be
U_{+1}(\tilde{a}) \equiv   {1 \over \tilde{a}^2}  
 \Big( \log{ {1 \over \tilde{a}^4 } } -1\Big) - \gamma^2 , ~~~\text{where} ~~~~~0 \leq \gamma^2 .
\ee
The  equation  $U_{+1}(\tilde{a}) =0$ defines  the maximal value of the scale factor $\tilde{a}  = \mu$ through the equation
\be\label{muk1}
  {1 \over \mu^2}  
 \Big( \log{ {1 \over \mu^4 } } -1\Big) - \gamma^2=0.~~~~
  \ee
The substitution 
\be
 \gamma^2 \mu^2 = 2 u
\ee
reduces it to the Lamber-Euler equation
\be
~~~~u e^{u} = {\gamma^2 \over 2 \sqrt{e}}.
\ee
The solution is expressible in terms of the $W_0(x)$ function, which is defined in the positive interval region $ 0 \leq x \leq \infty$:
\be
u = W_{0}({\gamma^2 \over 2 \sqrt{e}}),
\ee
and it allows to express the maximal value of the scale factor:
\be\label{amaxk1}
\mu^2 = {2\over \gamma^2} W_{0}({\gamma^2 \over 2 \sqrt{e}})~~~~\text{and 
}~~~~0 \leq \mu^2 \leq  {1 \over \sqrt{e}}.
\ee
Thus the scale factor $ \tilde{a}$ takes its values in the interval
\be
~~~\tilde{a} \in [0, \mu]
\ee
shown in Fig.\ref{figU1}. With the next substitution  
\be\label{substitute}
   \tilde{a}^4  =\mu^4 e^{-b^2},~~~~  ~ b \in  [-\infty,\infty].
\ee
the equation (\ref{k1case})  will take the following form:  
\be
{d b \over d \tau}  =   {2 \over \mu^2}  ~   e^{{b^2 \over 2}}   \Big(1    +     {\gamma^2  \mu^2 \over b^2}  (1-  e^{-{b^2 \over 2}}   \Big)^{1/2}.
\ee
 Integrating the equation with the boundary conditions at $\tau =0$ where $\tilde{a}(0) = 0$ and 
$b(0)=-\infty$ we will get the parametric representation of the function $b(\tau)$:
\beqa\label{solutionk1}
&&\int^{b(\tau)}_{-\infty}    { d b   e^{-{b^2 \over 2}}   \over 
  \Big(1    +      { \gamma^2  \mu^2 \over b^2}  (1-  e^{-{b^2 \over 2}}   \Big)^{1/2}        } ~ = {2 \over \mu^2}   ~ \tau.
\eeqa
\begin{table}[htbp]
   \centering
     \begin{tabular}{@{} cccccc @{}} 
     
      \toprule
      $\gamma^2$ &  $\mu^2  $ &  $2 g^2 \CF_{m}/\Lambda^4_{YM}$ & $ \epsilon_{m}/\Lambda^4_{YM}$  & $ p_{m}/\Lambda^4_{YM}$& $\tau_0$     
      \\
      $\gamma^2 =\Big({L \over a_0}\Big)^2$     &~ $\tilde{a }\in[0,\mu]$    &${ 1 \over \mu^4 }  $& $  \CA { \gamma^2  \over \mu^2 }      $&  $ {  \CA  \over 3 \mu^4 }   \Big(4+   \gamma^2 \mu^2 \Big)  $  & 
      \\  
      \midrule
      $\gamma^2 \rightarrow 0$                     &~~$1/ e^{1/2}$             & $ e  $         &0             &  ${ 4  e\CA  \over 3 }   $     &  $\sqrt{{\pi \over 8 e}}$ \\
 $\gamma^2 \rightarrow \infty$   & ${2\over \gamma^2}  \log\gamma^2 $  &    ${ \gamma^4 \over 4 \log^2\gamma^2 }    $              & $ {\CA \gamma^4 \over 2   \log\gamma^2}$        &${1\over 3} {\CA \gamma^4 \over 2   \log \gamma^2}+  {\CA \gamma^4 \over 3 \log^2\gamma^2 } $   & ${\sqrt{2\log\gamma^2} \over \gamma^2} $
  \\
      \bottomrule
   \end{tabular}
   \caption{The behaviour of the field strength tensor and of the components of the energy momentum tensor are shown for $k=1$ and  $\gamma^2 \in [0, \infty] $.   When $\gamma^2 \rightarrow 0$, the solution reduces to the case  $k=0$.   In the limit $\gamma^2 \rightarrow \infty$ the universe is contracting   to the zero size and physical observables are diverging as $2 g^2 \CF ={ \gamma^4 \over 4 \log^2{\gamma^2 }  }   \rightarrow \infty  $ and  with $ \epsilon$ and $p$ having a similar behaviour.  
   }
   \label{tbl:largeS}
\end{table}
From this equation we obtain $b(\tau)$ by inversion of this elliptic-type integral. 
The time interval $\tau \in [0,\tau_m]$ during which the scale factor is reaching its maximal value (it is equal to a half of the total period, where $b=0$) can be expressed through the integral 
\beqa
&&\int^{0}_{ -\infty}    { d b   e^{-{b^2 \over 2}}   \over 
  \Big(1    +       { \gamma^2  \mu^2  \over b^2}  (1-  e^{-{b^2 \over 2}}   \Big)^{1/2}        } ~ = {2 \over \mu^2}   ~ \tau_m.
\eeqa
The field strength (\ref{fieldstrength}) evolution in time is expressible in terms of $b(\tau)$  function:
\be
2 g^2 \CF 
=  { e^{b^2(\tau)}  \over \mu^4 }   \Lambda^4_{YM}. 
\ee
The energy density and pressure (\ref{energypressure3}) will evolve in time as well:
\beqa\label{enerpressk1}
 &&\epsilon  =  {  \CA  \over \mu^4 } e^{b^2(\tau)}  \Big( b^2(\tau)+ \gamma^2 \mu^2 \Big) \Lambda^4_{YM} ,~~~~~~~~
~~p =  {  \CA  \over 3 \mu^4 } e^{b^2(\tau)}  \Big(4+ b^2(\tau)+ \gamma^2 \mu^2 \Big) \Lambda^4_{YM}.  
 \eeqa
The Table \ref{tbl:largeS} summarises the data for the  $k=1$ case. The equation of state will take the following form: 
 \be\label{eqstak1}
 p = {\epsilon \over 3} +  ~{ 4 \CA  e^{b^2(\tau)}    \over 3 \mu^4 }  \Lambda^4_{YM} .
 \ee
The equation has an additional term that increases  the pressure.   The minimal value of the pressure and of the field strength tensor (\ref{fieldstrength}) is reached at  the midpoint $\tau =\tau_m$ where $b(\tau_m)=0$:  
\beqa\label{variation}
2 g^2 \CF_{m} =  { 1 \over \mu^4 }  \Lambda^4_{YM},~~~~~~ \epsilon_{m} =      {  \CA \gamma^2  \over \mu^2 }    \Lambda^4_{YM} ,~~~~~
p_{m} =   { \epsilon_{m}  \over 3}  +  {  4 \CA   \over 3 \mu^4 }   \Lambda^4_{YM}.
\eeqa
The parameter  $\mu^2 $ (\ref{amaxk1}) varies in the interval $0 \leq \mu^2 \leq  {1/e^{1/2}}$, therefore  $e \Lambda^4_{YM} \leq  2 g^2 \CF_{m} $ and $0 \leq \epsilon_{m}$. Only the positive part of the energy density curve is involved in the evolution.

Let us consider the limiting behaviour at $\gamma^2 \rightarrow 0$  and $\gamma^2 \rightarrow \infty$ (see Table \ref{tbl:largeS}). 
In the limit $\gamma^2 \rightarrow 0$  the solution (\ref{solutionk1}) reduces to the case $k=0$ that was already considered in previous section. In the 
second limit the maximal value of the scale factor (\ref{amaxk1}) tends to  zero $\mu^2  ={2\over \gamma^2}  \log{\gamma^2 }   \rightarrow 0$ and the whole universe contracts to the zero size and physical observables are therefore diverging $2 g^2 \CF ={ \gamma^4 \over 4 \log^2{\gamma^2 }  }   \rightarrow \infty  $, the  $ \epsilon$ and $p$ are diverging as well.   

For a typical  value of $\gamma^2$, let's say  $\gamma^2 =1$, we have   $ \mu^2 =  2 W_0(1/2\sqrt{e} ) \simeq 0.48$,  $\tau_m \simeq 0.27$, and the duration of the expansion is:
\be\label{k1per}
k=1,~~~~~~ c t_m= \tau_m L  \simeq  0.27  L ~   \simeq    1.12 \times 10^{14 }  \Big({eV \over \Lambda_{YM} }\Big)^2 ~ cm. 
\ee
Using (\ref{muk1}) and (\ref{enerpressk1}) for the deceleration parameter (\ref{deceleration}) one can get:
\be\label{de-acceleration1}
q={b^2 +   \gamma^2 \mu^2 + 2   \over b^2 +  \gamma^2 \mu^2 (1 -e^{-b^2/2} )   }
  \geq 1.
\ee
There is no acceleration at any time.  When $\gamma^2 \rightarrow 0$ this expression reduces to the $k=0$ expression (\ref{de-acceleration0}).   
The Hubble parameter  and the density parameter $\Omega$ (\ref{omega}) are:
\be
H^2 = {  e^{b^2} \over L^2 \mu^4} \Big( b^2 + \gamma^2 \mu^2  (1 -e^{-b^2/2} ) \Big)  ,~~~~\Omega_{vac} -1 =  {\gamma^2 \over ({d \tilde{a} \over d \tau})^2 } =  {\gamma^2 \mu^2 e^{-b^2/2} \over b^2  +  \gamma^2 \mu^2 (1 -e^{-b^2/2} ) }. 
\ee
As one can see, the general behaviour of the solutions in the cases considered in the last two sections: $k=0$, $k=1$ and Type I solution $k=-1$ at  $0 \leq \gamma^2  < {2\over \sqrt{e}}$ are qualitatively the same. They all describe a closed universe, as it is shown in Fig.\ref{figU0} for $\tilde{a}(\tau)$. In all these cases the initial value of the scale factor  is zero: $a(0)=a_0 \tilde{a}(0)=0$.  The corresponding half-time periods of the expansion are given in  (\ref{k0per}), (\ref{k1per}) and  (\ref{km1per}).  

\section{\it Primordial Gravitational Waves }
The coefficient of amplification of primordial gravitational waves obtained  by Grishchuk in \cite{Grishchuk} 
has the following form:
\be\label{amplification} 
K = {1\over 2} \Big(   {\beta \over  n \eta_0}  \Big)^2, ~~~~~\beta = {1-3 w \over 1+3w},\nn
\ee
where $w$ is  the  barotropic  parameter  (\ref{standradmatter}), $n$ is a wave number,  and the wavelength is $\lambda = 2\pi a/ n$. A gravitational wave is amplified when the equation of state  differs from the evolution $a \propto \eta$ dictated by the radiation ($p = {1\over 3}\epsilon$, $w= 1/3$), and earlier this wave had been initiated at $\eta_0$.  The  gravitons are produced with particularly great intensity during the initial exponential expansion of the universe where $\eta_0 \sim 0$. The production of gravitons is slowing down when the expansion takes on a form characteristic to a hot universe ($w=1/3$). The behaviour of $K$ is singular when  $\eta_0 \rightarrow 0$ or when $n \rightarrow 0$.

\begin{figure}
 \centering
\includegraphics[angle=0,width=8cm]{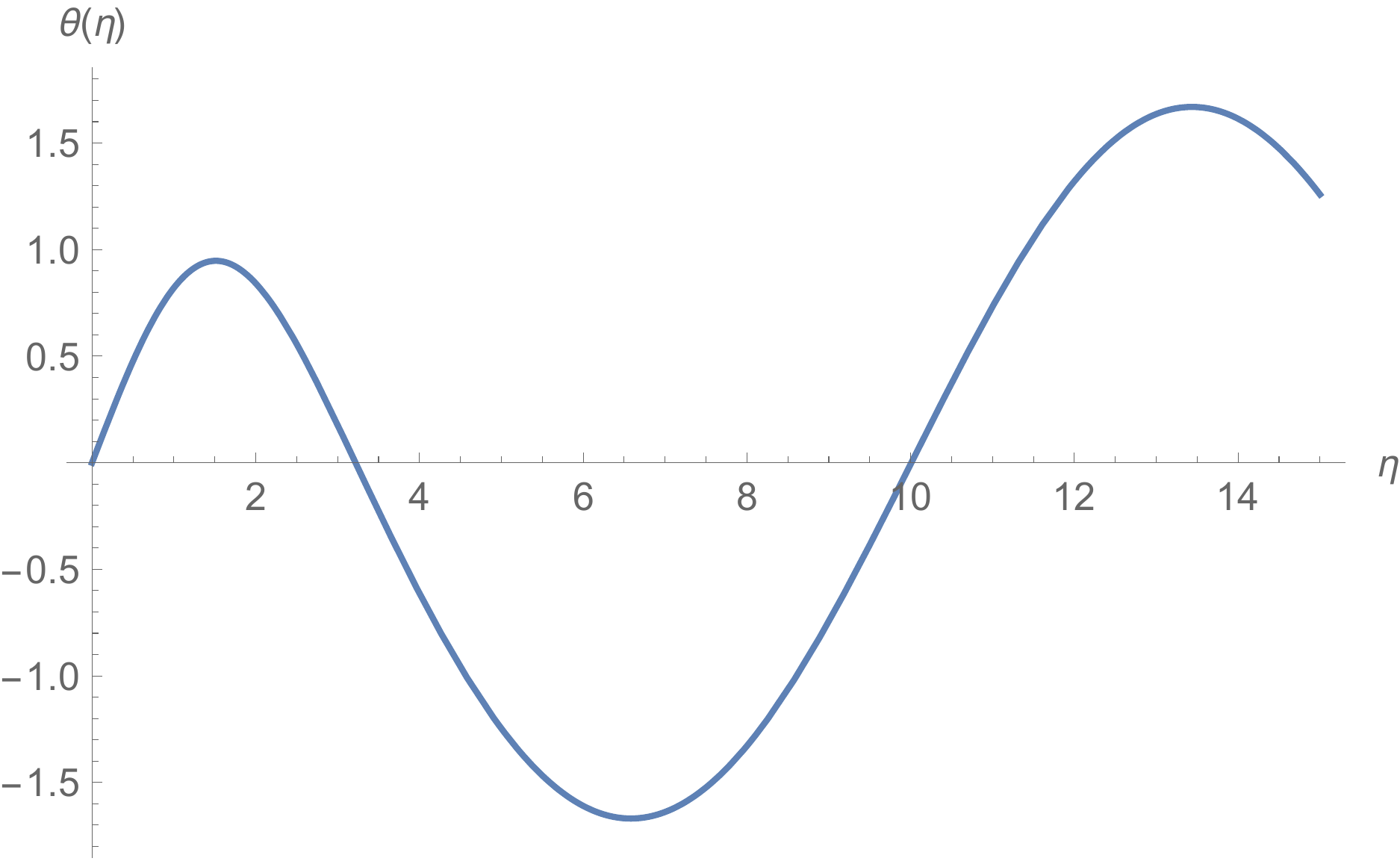}~~~~~
\centering
\caption{The perturbation of the Type II solution.  Here $k=-1$, $\gamma^2  \simeq 1.211$,   $\mu_2 \simeq 1.32$, the wave number $n = 1.01$ and $\tilde{a}(0)=\mu_2$,  $\theta(0)=0$, $\theta^{'}(0)=0.1$ in the equations (\ref{evolutionineta}) and (\ref{linearperturbationinYM}). }
\label{ampliYM} 
\end{figure}

The deviation from the radiation equation of state $w=1/3$ constitutes one of the necessary conditions for the amplification of primordial gravitational waves  \cite{Grishchuk}. As mentioned earlier, in the gauge field theory  there is a straightforward relation between  energy density and the pressure (\ref{relation}) of the form
 \be
 p = {1\over 3}\epsilon + {4\over 3}  {\CA \over \tilde{a}^4(\tau)} \Lambda^4_{YM} ~
  \ee
that genuinely deviates from the radiation equation of state $p = {1\over 3}\epsilon$ by an additional term, which depends on the beta function coefficient $\CA$  (\ref{mattergroupparameter}), the scale $\Lambda_{YM}$ and a time-dependent scale factor $\tilde{a}(\tau)$. The deviation is large at the initial stages of the universe expansion when $\tilde{a}(\tau) \rightarrow 0$ and tends to zero at late time when  $\tilde{a}(\tau) \gg \mu_2$. This takes place in the case of Type II and Type IV solutions.    

The equation describing the tensor perturbation $h_{\mu\nu}$ of the Friedmann space-time metric $\gamma_{\mu\nu}$  is of the form $g_{\mu\nu}= \gamma_{\mu\nu} + h_{\mu\nu}$ and has the following nonzero spacial components:
 \be\label{linearpert}
 h^i_j= h(\eta) ~Y^i_j ~e^{i n x} = {\theta(\eta)\over a(\eta)} ~Y^i_j ~e^{i n x},
 \ee
where $Y^i_j$ is the tensor eigenfunction of the Laplace operator. In conformal time $\eta$ and t-time the evolution of  the linear perturbation has the following form  \cite{Lifshitz:1945du}:
\be\label{linearpert1}
h^{''} + 2  {a^{'} \over a} h^{'}   + n^2 h  =0,~~~~~\ddot{h} +3 {\dot{a} \over a} \dot{h}  + {n^2 \over a^2} h=0,
 \ee
where $n $ is a wave number and the wavelength is $\lambda = 2\pi a/ n$.  The equation  (\ref{linearpert1}) for the $\theta$ amplitude in (\ref{linearpert}) reduces to the form  \cite{Grishchuk}:
\be\label{linearpert2}
\theta^{''} + \theta (n^2 - {a^{''} \over a})=0,
\ee
where the derivatives are over conformal time $\eta$ (\ref{conformaltime}), (\ref{basiceq1}).  For the  general parametrisation of the equation of state $p = w \epsilon$   the solution of Friedmann equations (\ref{FriedmannEquations}) and  (\ref{edens}) is:
\be\label{standradmatter1}
 ~~~\epsilon \ a^{3(1+w)} = const,~~~~~~a \sim t^{{2\over 3(1+w)}},~~~~~~a \sim \eta^{{2\over 1+3w}}~~~ \text{when}~~~ k=0.
\ee
Considering a universe with a "break"  at $\eta =\eta_0$ so that $a(\eta) = const $ and ${a^{''} \over a} =0$ for $\eta < \eta_0$,  Grishchuk  effectively introduced a potential barrier at $\eta =\eta_0$ into the equation (\ref{linearpert2}).   Substituting the solution (\ref{standradmatter1}) and the "break"  into the (\ref{linearpert2}) one can get   \cite{Grishchuk}
\be\label{Grishchuk}
\theta^{''} + \theta \Big(n^2 - \begin{cases}  {2(1-3 w) \over (1+3w)^2}{1\over \eta^2},~~~~\eta \geq \eta_0\\        0,~~~~~~~~~~~~~~~\eta < \eta_0\end{cases}    \Bigg\}  \Big)  =0.
\ee
With the potential barrier in place the amplification of waves (tensor perturbation) takes place when $n \eta_0 \ll 1$ and the  amplification parameter $K$  is given in (\ref{amplification}).  

Let us now consider the tensor perturbation of Type II solution of the Friedmann equations when the contribution (\ref{energymomentumYM01}) of the gauge field theory vacuum to the energy density of the universe is taken into consideration. For that consider the first first Friedmann equation  (\ref{basiceq1}) 
\be
\Big({\tilde{a}^{'} \over \tilde{a}  }\Big)^2=   {1\over \gamma^2} {1\over \tilde{a}^2 } \Big(\ln {1\over \tilde{a}^4} -1\Big)  -k  
\ee
together with the acceleration equation (\ref{accel}), which has the following form:
\be
{\tilde{a}^{''} \over \tilde{a}  } -  \Big({\tilde{a}^{'} \over \tilde{a}  }\Big)^2= -{1\over \gamma^2} {1\over \tilde{a}^2 } \Big(\ln {1\over \tilde{a}^4} +1\Big).
\ee
Adding together the last two equations gives
\be\label{evolutionineta}
\tilde{a}^{''}  = -{2\over \gamma^2} {1\over \tilde{a} } -k  \tilde{a} 
\ee
and the linear perturbation equation (\ref{linearpert2}) will take the form
\be\label{linearperturbationinYM}
\theta^{''} + \theta \Big(n^2 +{2\over \gamma^2} {1\over \tilde{a}^2 } +k \Big)=0.
\ee
In the case of the Type II solution where $\tilde{a}(0) = \mu_2$ (\ref{solak-1-II}) the system avoids a singular  behaviour in vicinity of $\eta=0$.  The amplification of the primordial gravitational waves is due to the second term in (\ref{linearperturbationinYM}) when $n^2 < {2/\gamma^2 \mu^2_2} $.  The  Fig.\ref{ampliYM} shows the behaviour of the linear perturbation  of the Type II solution. The analysis of the system of equations (\ref{evolutionineta}) and (\ref{linearperturbationinYM}) in details will be published elsewhere.

\section{\it Acknowledgement}

I would like to thank Viatcheslav  Mukhanov and Dieter L\"ust  for the invitation to the Sommerfeld Theory Colloquium  on the physics of the QCD vacuum \cite{Munich}, the kind hospitality at the LMU of Munich and the discussions that reiterate my interest to physics of the neutron stars \cite{Yerevan}, gravity \cite{body} and inflationary cosmology.  It is a pleasure to express my thanks to Luis Alvarez-Gaume for kind hospitality at the Simons Centre for Geometry and Physics where part of this calculation was initiated. This work was partially supported by the Science Committee of the Republic of Armenia, the research project No. 20TTWS-1C035.

\newpage

\section{\it Appendix A} 
\begin{figure}
 \centering
\includegraphics[angle=0,width=11cm]{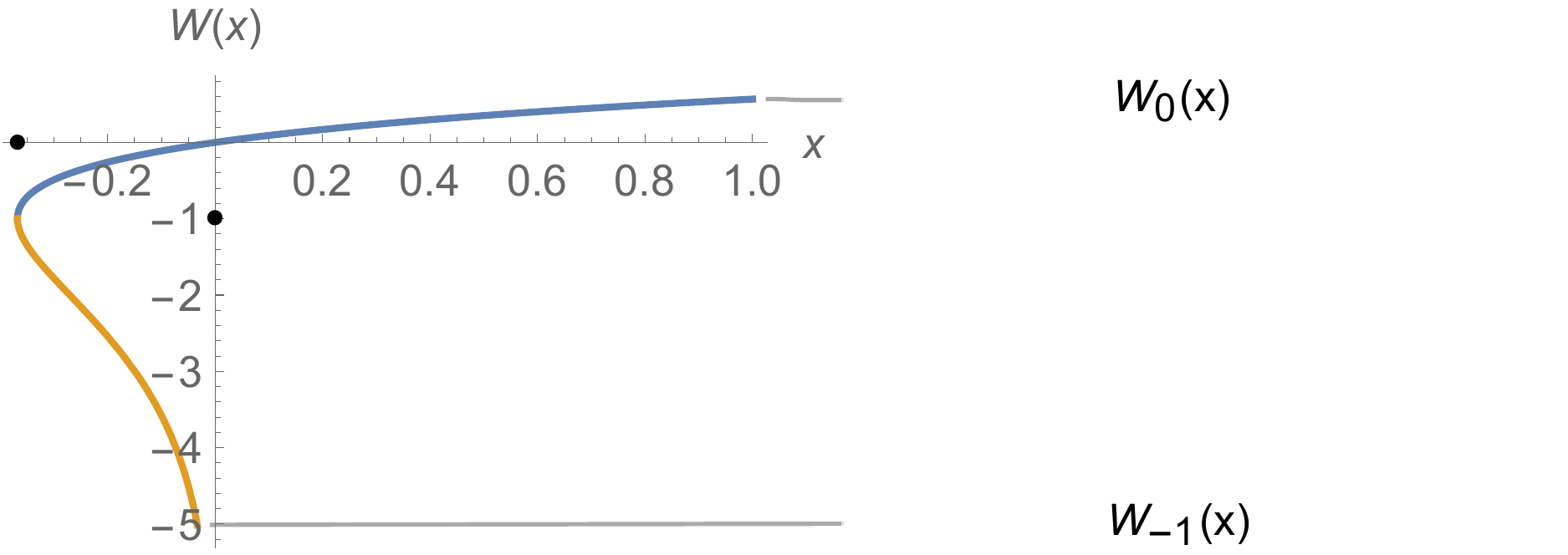}~~~~~
\centering
\caption{  The graph of W-function with its two real branches $W_0(x)$ and $W_{-1}(x)$. The two branches merge at the point $(-1/e,-1)$.
}
\label{Wfunction}
\end{figure}
The Lambert-Euler  W-function $W(x)$ is the solution of the equation \cite{Lambert,Euler,Corless}:
\be
W e^W = x.
\ee
There are two real branches of $W(x)$ (see Fig.\ref{Wfunction}). The solution for which $  -1 \leq  W(x) $ is the principal branch and denoted as $W_0(x)$. The solution satisfying $W(x) \leq -1$ is denoted by $W_{-1}(x)$. On the x-interval $[0,\infty)$  there is one real solution, and it is nonnegative and increasing. On the x-interval $[-1/e,0)$  there are two real solutions, one increasing and the other one decreasing.  Properties include:
\beqa
&&W_0(-1/e) = W_{-1}(-1/e) =-1,~~~ W_0(0)=0,~~~ W_0(e)=1,,~~~ W_0(e^{1+e})=e, \\
&&W_0(x)=\sum^{\infty}_{n=1} {(-n)^{n-1}   \over n!}x^n = x -x^2 +{3\over 2}x^3...,~~~ \vert x \vert < 1/e,~~~~~\nn\\
&&W_0(x) = \log x -\log\log x +\CO\Big({\log\log x \over \log x}\Big),~~~x \rightarrow +\infty\nn\\
&&W_{-1}(x) = -\log(-{1\over x}) -\log\log(-{1\over x}) +\CO\Big({\log\log(-{1\over x})\over \log(-{1\over x})}\Big),~~~x \rightarrow -0\nn
\eeqa

\end{document}